\documentclass[]{emulateapj}
\usepackage{natbib}
\usepackage{float}
\usepackage{multirow}
\usepackage{cases}

\newcommand{\kms}{km$\rm s^{-1}$}

\newcommand{\teff}{${T}_{\rm eff}$}
\newcommand{\logg}{log$g$}
\newcommand{\feh}{[Fe/H]}
\newcommand{\Mgb}{Mg$_b$}
\newcommand{\vR}{$\langle v_R\rangle$}
\newcommand{\vZ}{$\langle v_z\rangle$}
\newcommand{\vPhi}{$\langle v_\phi\rangle$}
\newcommand{\dvR}{$\Delta\langle v_R\rangle$}
\newcommand{\dvPhi}{$\Delta\langle v_\phi\rangle$}
\newcommand{\VPhi}{$v_{\phi,0}$}

\newcommand {\Msun}{{M$_{\odot}$}}
\newcommand{\ikpc}{kpc$^{-1}$}

\shorttitle{The peculiar velocities in the Galactic outer disk}
\shortauthors{HJT, CL, et al.}

\begin{document}

\title{The peculiar velocities in the Galactic outer disk---hints of the elliptical disk and the perturbation of the spiral structures}

\author{
Hai-Jun Tian\altaffilmark{1,2}, 
Chao Liu\altaffilmark{3}, 
Jun-Chen Wan\altaffilmark{3},
Li-Cai Deng\altaffilmark{3},
Zi-Huang Cao\altaffilmark{3},
Yong-Hui Hou\altaffilmark{4},
Yue-Fei Wang\altaffilmark{4},
Yue Wu\altaffilmark{3},
Yong Zhang\altaffilmark{4},
Yong-Heng Zhao\altaffilmark{3}
}

\altaffiltext{1}{Max Planck Institute for Astronomy, K\"onigstuhl 17, D-69117 Heidelberg, Germany; hjtian@lamost.org}
\altaffiltext{2}{China Three Gorges University, Yichang 443002, China}
\altaffiltext{3}{Key Lab for Optical Astronomy, National Astronomical Observatories, Chinese Academy of Sciences, Beijing 100012, China; liuchao@nao.cas.cn}
\altaffiltext{4}{Nanjing Institute of Astronomical Optics \& Technology, National Astronomical Observatories, Chinese Academy of Sciences, Nanjing 210042, China}

\begin{abstract}
We present the peculiar in-plane velocities derived from the LAMOST red clump stars. From the variations of the in-plane velocity with the Galactocentric radius for the young and old red clump stars, we are able to identify two types of peculiar velocities: 1) both the two red clump populations show that the radial velocity is negative within $R=9.0$\,kpc and becomes positive beyond (denoted as the \emph{long-wave} mode); and 2) the young red clump stars show larger mean radial velocity than the old population by about 3\,\kms\ between $R=9$ and 12\,kpc (denoted as the \emph{short-wave} mode). We find that the elliptical disk induced by the rotating bar can well explain the \emph{long-wave} mode peculiar velocity. The axis ratio of the elliptical disk is around 0.8-0.95 and the disk keeps circular at $R=9.24\pm0.2$\,kpc, which should be the location of the outer Lindblad resonance radius (OLR). Adopting the circular speed of 238\,\kms, the pattern speed of the bar is then derived as $48\pm3$\,\kms\ikpc\ from the location of OLR. On the other hand, the \emph{short-wave} mode is likely the perturbation of the spiral arms as density waves. 
\end{abstract}
\keywords{Galaxy: disk --- Galaxy: structure --- Galaxy: kinematics and dynamics ---Galaxy: stellar content --- stars: kinematics and dynamics}

\section{Introduction}\label{sect:intro}

In recent years, the Galactic disk is found not symmetric in both the star density and kinematics. \citet{widrow2012} claimed that the vertical stellar density shows small-amplitude but significant oscillation in both the north and south of the Galactic mid-plane. \citet{xu2015} further found the wave-like ring in the outer disk. \citet{siebert2011a} discovered from the RAVE~\citep{siebert2011b} survey data that the radial velocity in the solar neighborhood is not symmetric, the stars show about 7 \kms\ toward the inner Galaxy. \citet{bond2010} also confirmed this from the SDSS survey data. \citet{carlin2013}, \citet{williams2013}, and \citet{sun2015} demonstrated that the asymmetric motion is not only in radial direction, but also in vertical direction. And \citet{tian2015} found that the asymmetric motion may be related to the age of the stars. \citet{bovy2015} studied the power spectrum of the velocity map in the $X$--$Y$ plane for the APOGEE~\citep{alam2015} red clump stars and argued that the velocity oscillation with a broad peak at the wave number of 0.2--0.9\,\ikpc\ is likely due to the Galactic central bar.

Several theoretical works provided possible explanations about how the asymmetric stellar density and oscillated velocity happen. \citet{gmo+13} thought that the asymmetric vertical stellar density and the bulk motions can be the result of the perturbation of the minor merger. \citet{siebert2012}, who used a simplified model of the perturbation from the spiral density wave and~\citet{faure2014} ran N-body simulations with spiral arms to explain that the asymmetric motion revealed by \citet{siebert2011a} could be the induced by the spiral structures. \citet{debattista2014} found from the N-body simulations that the (strong) spiral structures can induce vertical oscillation near the disk mid-plane. Recently, \citet{grand2015} applied the power spectrum approach to the N-body simulation data and found that the bar with transient co-rotating spiral structures can produce similar features found in \citet{bovy2015}.

Theoretically, the peculiar velocities can be either induced by the perturbation of the non-axisymmetric structures or can geometrically reflect the lopsidedness of the disk. Many works have discussed about the former scenario \citep{siebert2012,gmo+13,debattista2014}. Here we highlight more about the latter.

Lopsidedness, which means the surface brightness of a galactic disk is not perfectly axisymmetric but varies with the azimuthal angle in the polar coordinates, is a common feature in galaxies~\citep[][etc.]{richter1994,rix1995,zaritsky1997,swaters1999,conselice2000,bournaud2005}. \citet{rix1995} decomposed the surface brightness of 18 face-on galaxies to Fourier series in polar coordinates and found many of them show significantly large amplitudes from mode 1 to mode 6. Particularly, the bar and two-arm spiral structure can naturally form the mode 2 structure, i.e. the disk looks elliptic rather than circular. The lopsided and elliptic disk should be in state status and thus the stars in these non-axisymmetric disks must have coherent peculiar motion biased from the axisymmetric velocities. \citet{kuijken1994} and \citet{rix1995} derived the features of both the geometry and the coherent radial and azimuthal velocities for the disk given a non-rotating elliptical gravitational potential. Specifically, the major axis of the elliptical disk is perpendicular to the major axis of the elliptical gravitational potential. The ellipticity of the surface brightness within the underlying elliptical potential smoothly disappear when the radius become large \citep[see Eq. (11) in][]{rix1995}. Meanwhile, the maximal azimuthal velocity occurs at the points of minor axis, while the maximum of radial velocity shift to 45$^\circ$ from the minor axis. Based on the non-rotating elliptical gravitational potential, \citet{kuijken1994} investigated the ellipticity of the disk of the Milky Way and found the axis ratio of the iso-potential is about 0.9.

However, these models for the elliptical disk do not account for the rotation of the non-axisymmetric component. For a disk with a rotating bar in the central region, \citet{binney2008} derived the radius of the closed loop orbit as the function of the azimuth angle (see Eqs (4.148a,b)) and displayed that the major axes of the closed loop orbits change their direction at various radii: inside the Inner Lindblad Resonance radius (ILR) the major axis of the closed loop orbits are aligned with the minor-axis of the bar; between the ILR and the Corotation Resonance radius (CR) they are aligned with the major-axis of the bar; between the CR and the Outer Lindblad Resonance radius (OLR) they are aligned with the minor-axis again; and beyond the OLR they are aligned back with the major-axis of the bar. Therefore, the ellipticity of the disk with a central rotating bar is much more complicated than the models introduced in \citet{kuijken1994} and \citet{rix1995} in the sense that not only the axis ratio may changes with radius, the phase of the ellipticity also changes with radius.


In this paper, we use the red clump stars, which are mostly distributed in the Galactic anti-center region, from the LAMOST survey~\citep{Cui2012,Zhao2012} to investigate the peculiar in-plane velocities from Galactocentric radius of $R\sim7$ to 14\,kpc in the Galactic disk. We identify the peculiar velocity features from both the ellipticity of the bar and from the local perturbation of the spiral structures from our sample.

The paper is organized as follows. In Section 2, we describe how to select the proper red clump stars from the LAMOST catalog and separate them into young and old population. In Sections 3, we discuss the approach to derive the in-plane velocity based on the velocity de-projection technique. In Section 4, we demonstrate our results about two kinds of peculiar in-plane velocities, the \emph{long-wave} and \emph{short-wave} oscillations.  In Section 5, we discuss the possible mechanisms which induce these peculiar velocities. Finally, we draw brief conclusions in Section 6.

\section{Data}\label{sect:data}

Large Aperture Multi-Object fiber Spectroscopic Telescope (LAMOST; also called Guo Shoujing telescope), is a quasi-meridian reflecting Schmidt telescope with the effective aperture of about 4 meter. In its 5$^\circ$ field of view,  4000 fibers are installed, which is capable of obtaining the same number of low resolution spectra (R$\sim1800$) covering from 380 to 900\,nm simultaneously~\citep{Cui2012,Zhao2012}. The on-going LAMOST spectroscopic survey is focusing on both the stars and some bright extragalactic objects. In its 5-year survey plan, it will obtain a few millions of stellar spectra in about 20,000 square degrees in the northern hemisphere~\citep{Deng2012}.

As December 2015, LAMOST has delivered its second data release (DR2), containing about 4 million stellar spectra. Among them, the LAMOST pipeline has provided stellar astrophysical parameters (effective temperature, surface gravity, and metallicity) as well as line-of-sight velocities for about 2.2 millions FGK type stars~\citep{wu2014}. Most of the spectra have well measured line indices~\citep{liu2015a}. Based on the line indices and applying the classification approach developed by~\citet{liu2014b}, about half million K giant stars are identified from LAMOST DR2. \citet{liu2015b} has improved the surface gravity (\logg) estimation with uncertainty of $\sim0.1$\,dex for the metal-rich giant stars. \citet[][]{wan2015} subsequently identified about 120 thousands of red clump (RC) star candidates from LAMOST DR2 catalog based on the empirical distribution model in the \teff--\logg--\feh\ space. Not only the primary RC stars, which are usually considered as standard candles, but also the secondary RC stars, which are more massive and young helium core burning stars, are identified in~\citet{wan2015}. The distances of both types of RC stars are estimated by them from the isochrone fitting with accuracy of $\sim10$\%.

\subsection{Remove the RGB contaminations}\label{sect:RGBcont}

\begin{figure}[!t]
\centering
\includegraphics[width=0.45\textwidth, trim=0.0cm 0.0cm 0.0cm 0.0cm, clip]{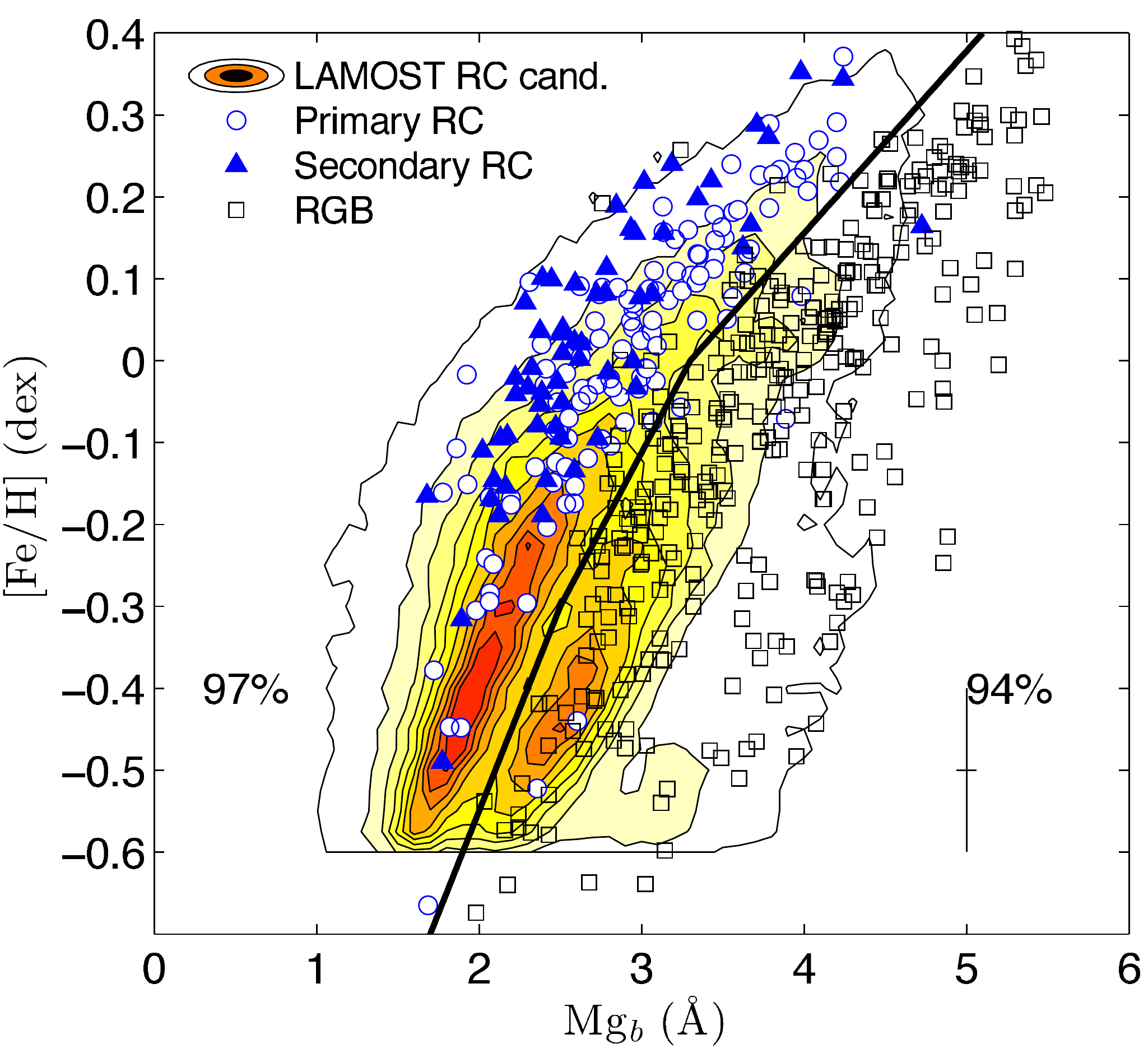}
\caption{The distribution of the RC candidate stars from~\citet{wan2015} in the \feh-\Mgb\ plane. The contours show the number density of the LAMOST RC candidates. The black rectangles, blue circles, and blue filled triangles are the red giant branch stars, the primary red clump stars, and the secondary red clump stars identified by~\citet{stello2013}, respectively.}
   \label{fig:MgFeH}
\end{figure}

The technique used in~\citet{wan2015} is based on statistics, i.e., for each star, the method can give the probability to be a RC star. A small fraction of red giant branch (RGB) stars are then mixed in the RC candidate catalog. In order to remove these contaminations, we apply another technique mentioned in the Appendix of \citet{liu2014b} to distinguish the RGB stars from the RC candidates. We map the RC candidates to the \feh\ vs. \Mgb\ plane, where \Mgb\ is the Lick index for the spectral line of Mg I located around 5184\,$\rm\AA$ based on~\citet{worthey1994}, in Figure~\ref{fig:MgFeH}. It is obviously seen that the density map of the RC candidates in Figure~\ref{fig:MgFeH} shows clear bimodality with a elongated valley from \feh=-0.6\,dex and \Mgb$\sim2$\,$\rm\AA$ to \feh$\sim-0.1$\,dex and \Mgb$\sim2.5$\,$\rm\AA$. Because the RC stars are always warmer than the RGB stars given the same metallicity and surface gravity, they should have relatively smaller values of \Mgb\ than the RGB stars with same \feh\ and \logg. On the other hand, \logg\ of the RC stars should be smaller than that of the RGB stars with same metallicity and effective temperature. The smaller \logg\ also leads to smaller value of \Mgb. Combining the two trends together, given a metallicity, the RC stars always locate at the side with smaller \Mgb\ than the RGB stars. 

Therefore, the clump with smaller \Mgb\ should be contributed by the RC stars, while the another clump with larger \Mgb\ should be contributed by the RGB stars. For the stars with \feh$>-0.1$, the gap between the RC and the RGB stars is not quite clear. We then overlap the common stars between the LAMOST DR2 and \citet{stello2013}, who classified the RGB, primary and secondary RC stars from the seismic features in the figure. The RC stars (blue hollow circles for the primary RC stars and the blue filled triangles for the secondary RC stars) and the RGB stars (represented with the black hollow rectangles) from~\citet{stello2013} are well separated in the \feh--\Mgb\ plane. Then, we are able to extend the separation between RGB and RC stars from metallicity lower than $-0.1$\,dex to super-solar metallicity using the empirical separation line from (\Mgb, \feh)$=$(2.1, -0.5), (2.5, -0.3), and (3.3, 0.0) to (5.0\,$\rm\AA$, 0.38\,dex) (the thick black line). 

Testing the performance of the empirical separation, we find 97\% seismic-identified RC stars are also identified as the RC stars from the separation line and 94\% seismic-identified RGB stars are also identified as RGB stars. If the RC candidate catalog from~\citet{wan2015} suffers 30\% RGB contaminations, the empirical separation shown in Figure~\ref{fig:MgFeH} can reduce the contamination by only about 2\%, according to the data from~\citet{stello2013}.

\subsection{The young and old RC populations}\label{sect:groupRC}

\begin{figure*}[!t]
\centering
\includegraphics[width=0.95\textwidth, trim=0.0cm 0.0cm 0.0cm 0.0cm, clip]{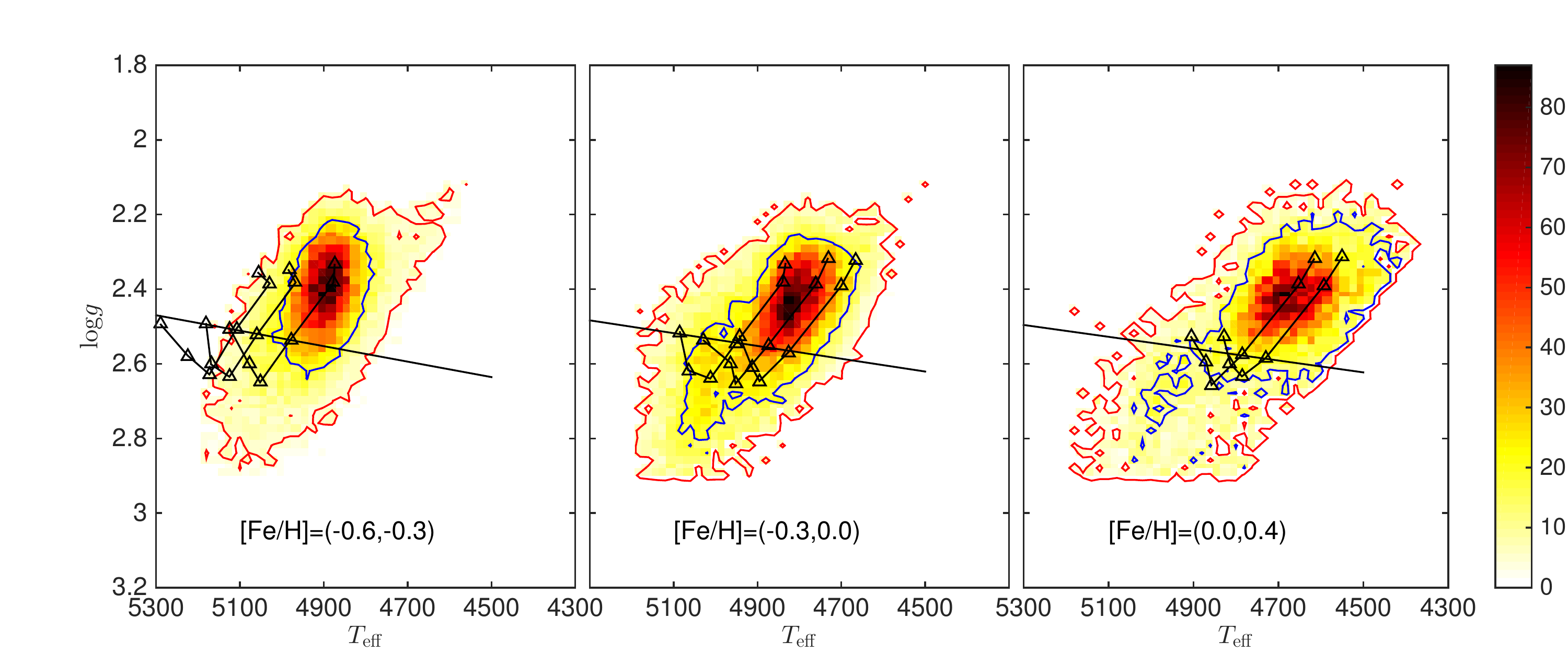}
\caption{The separation of the young and old RC stars in three \feh\ bins, -0.6$<$[Fe/H]$<$-0.3 (left panel), -0.3$<$[Fe/H]$<$0 (middle panel), and 0$<$[Fe/H]$<$0.4 (right panel). The black triangles are the isochrones for the helium core burning stars from~\citet{bressan2012}. From the left to the right in the left panel, the isochrones are at \feh$=$-0.6, -0.5, and -0.4. For each isochrone, the triangles present for 0.2, 0.5, 1.0, 2.0, 3.0, and 5.0\, Gyr from left to top-right. In the middle panel, the isochrone tracks are at \feh$=$-0.3, -0.2, and -0.1 from left to right, respectively. In the right panels, the isochrones are at \feh$=$0.1 and 0.3 from left to right, respectively. The straight lines are the best fit separation lines at 2\,Gyr in different ranges of \feh. Below the lines are the isochrone-based young population with age$<2$\,Gyr and above are the old population with age$>2$\,Gyr.}
   \label{fig:divide2}
\end{figure*}

The primary RC stars are those in the helium core burning stage which had the degenerate helium core in late RGB stage and have been triggered the helium flash. On the contrary, the second RC stars have initial stellar mass larger than the critical mass of helium flash (around 2\Msun) and thus ignite the helium core before the core become electronic degenerate. Because the different evolutionary tracks, the positions for the two types of stars in the Hertzprung--Russell diagram (or equivalently in the \teff--\logg\ diagram) are slightly different. Moreover, because the initial stellar mass for the secondary RC stars are larger, they are in general younger than the primary RC stars, which is helpful to trace the evolution of the Galactic disk. However, in practice, it is very difficult to separate the two kinds of stars directly from the \teff--\logg\ diagram. Although they can be separated from the their distinct asteroseismic features~\citep{bedding2011,stello2013}, only very few stars have accurate seismic measurements. Therefore, we turn to separate the young and old populations of RC stars with the help of the isochrones. We use the PARSEC isochrones~\citep{bressan2012} for the age separation. The triangles shown in the panels of Figure~\ref{fig:divide2} are the theoretical positions of the helium core burning stars in \teff--\logg\ plane with different metallicities (from the left to the right panels are \feh=(-0.6, -0.3), (-0.3,0.0), and (0.0, 0.4), respectively) . From the isochrones in each \feh\ bin, we select a straight line best fitting the data points at 2\,Gyr as the separation line. The RC stars younger than 2\,Gyr are mostly located below the lines. Although the very young and massive Helium core burning stars ($>10$\,$M_\odot$) can go above the lines, they should be very rare and may not affect our later analysis. Therefore we ignore these stars and define all RC stars located below the 2\,Gyr line are the young population and the RC stars located above the 2\,Gyr lines are old.

\begin{figure}[!t]
\centering
\includegraphics[scale=0.5]{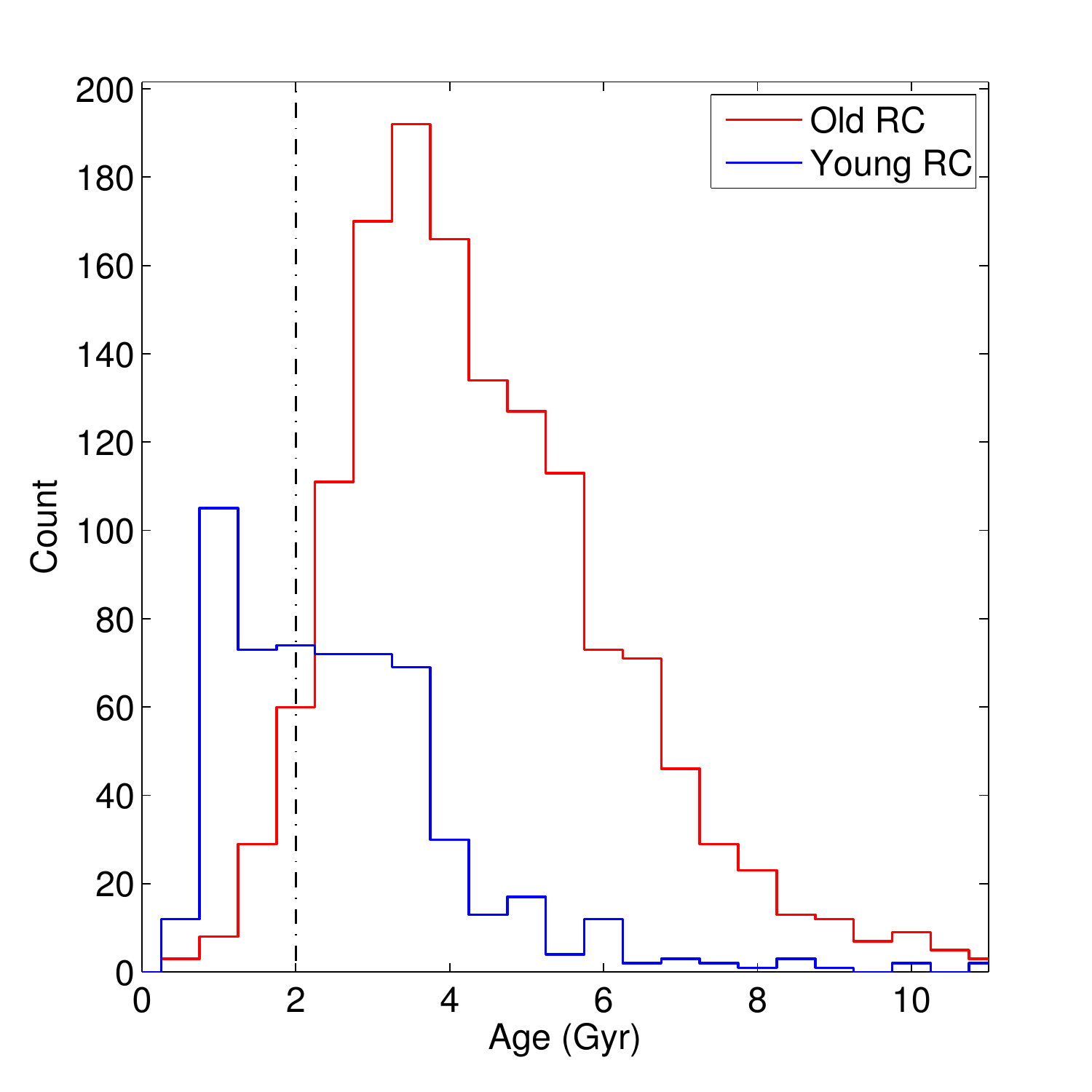}
\caption{The distribution of the age for the isochrone-separated young (blue line) and old (red line) RC stars. The age is derived from [C/N] by~\citet{martig2016}. The vertical black dot-dashed line indicates the position of 2\,Gyr.}\label{fig:agehist}
\end{figure}

In order to assess whether the isochrone-based age separation is feasible, we cross-match the LAMOST RC catalog with the APOGEE data with age estimates from the [C/N] by~\citet{martig2016}. We find about 2100 common RC stars after cross-matching. Figure~\ref{fig:agehist} shows the distributions of the age derived by~\citet{martig2016} for the young (blue line) and old (red line) RC stars identified from Figure~\ref{fig:divide2}. We find that the young RC stars are indeed dominated by young stars, although there is also many samples located beyond 2\,Gyr. And most of the old RC stars are located in the right side of the 2\,Gyr line (the vertical black dotted-dash line) in the figure. The mean values (standard deviations) of the age are 2.7 (1.6) and 4.6 (2.1)\,Gyr for the isochrone-identified young and old RC stars, respectively. Note that the age derived from~\citet{martig2016} may not be consistent with the PARSEC isochrones, it could have systematical difference in age, which can explain why the young RC stars have larger age in~\citet{martig2016} than what the isochrones expect. However, in this work, we do not need accurate age separation but only mark the young and old populations and focus on the comparison of the stellar kinematics. For this purpose, the isochrone-based young/old population separation is sufficiently helpful.

\subsection{The final sample}\label{sect:finaldata}

\begin{figure*}[!t]
\centering
\includegraphics[width=0.95\textwidth, trim=0.0cm 0.0cm 0.0cm 0.0cm, clip]{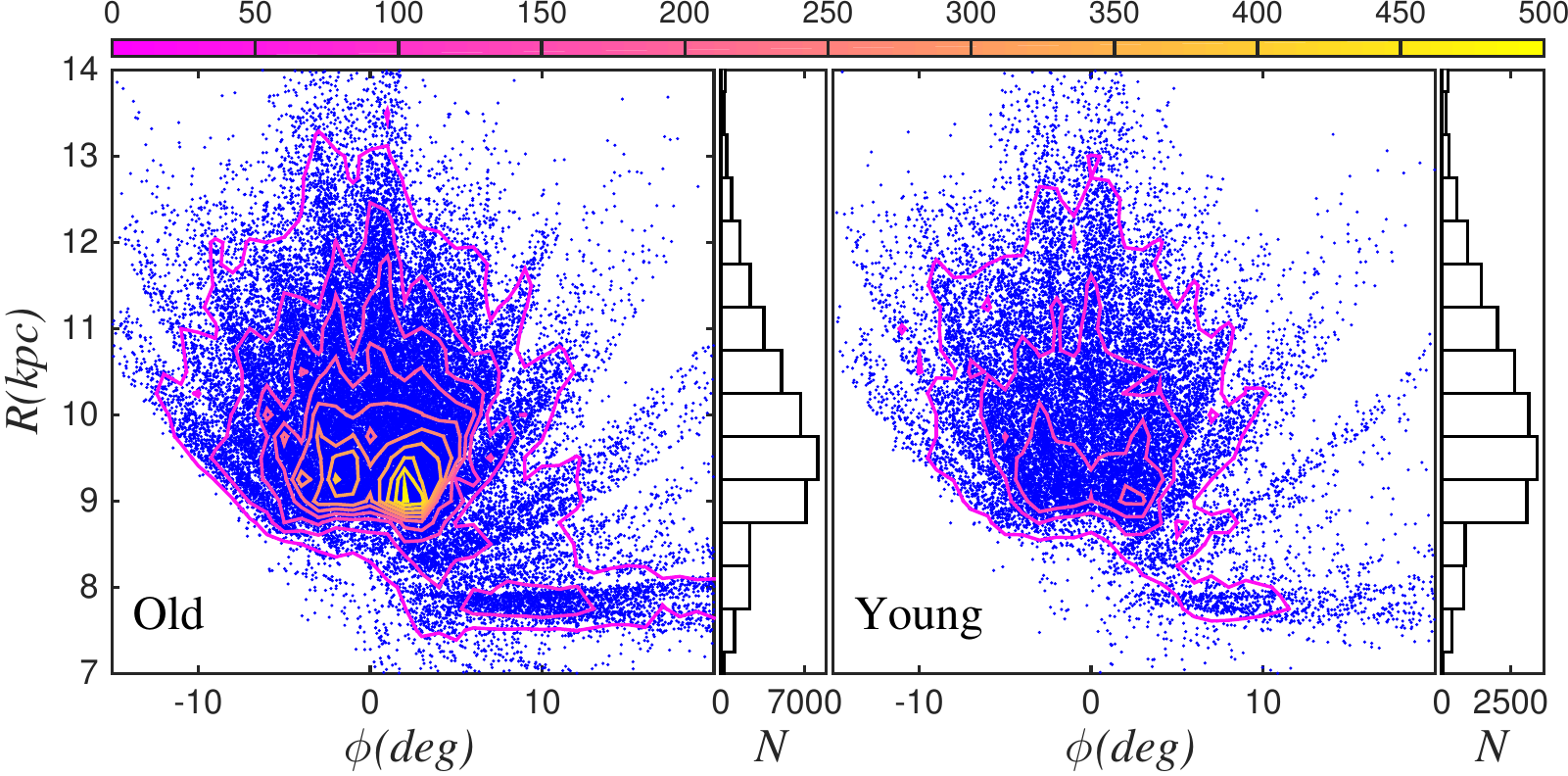}
\includegraphics[width=0.95\textwidth, trim=0.0cm 0.0cm 0.0cm 0.0cm, clip]{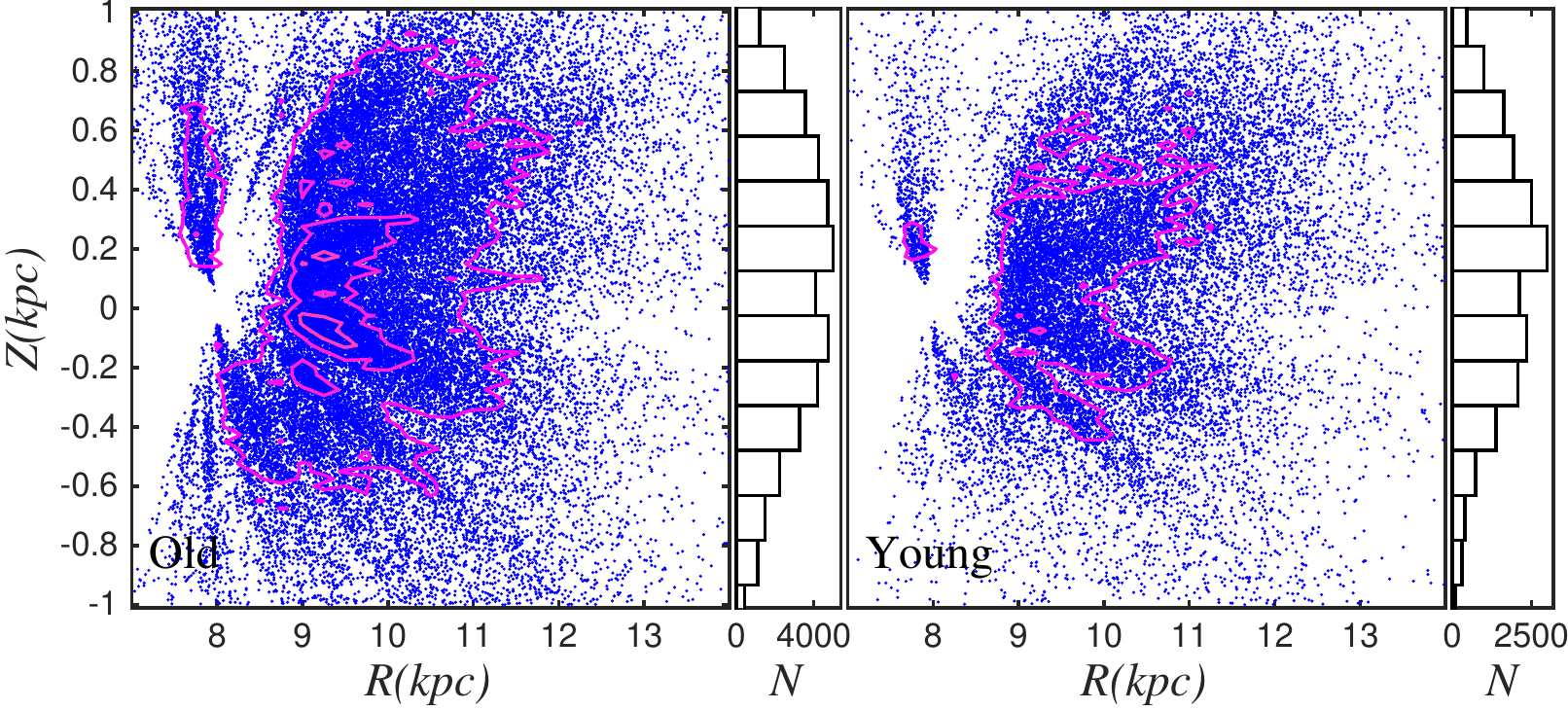}
\caption{The spatial density distribution, in Galactic cylindrical coordinates, for the old (left column) and young (right column) RC populations. The color contours represent the number density of the stars with levels of from 50 to 500 stars per bin with step of 50. }
   \label{fig:density_distri}
\end{figure*}

We further select the RC stars close to the Galactic mid-plane with the criterion of $|z|<1000$\,pc, where $z$ is the vertical height above or beneath the mid-plane for a RC star. And to ensure that the stellar parameters applied in the above data selection is reliable, we only select the stars with signal-to-noise ratio larger than 10. Finally, we obtain 42,713 old and 19,963 young RC stars. Figure~\ref{fig:density_distri} displays the spatial distribution of the old (in left panels) and young (in the right panels) RC stars in the three dimensional Galactocentric cylindrical coordinates. The top row is for the $R$--$\phi$ plane and the bottom row is for the $R$--$z$ plane. The smoothed color contours represent for the number density of the stars.

\section{The method}\label{sect:method}
Ideally, to obtain the in-plane velocity of the Galactic disk, we need to know the tangential velocity as well as the line-of-sight velocity for the tracers. However, as shown in Figure~\ref{fig:density_distri}, lots of the samples are too far to have reliable proper motion measures. Therefore, we cannot directly obtain the tangential velocities. However, if we assume that the mean velocity components, namely \vR, \vPhi, and \vZ, are same at given Galactocentric radius $R$ for the dataset, we can derive them from the velocity de-projection technique~\citep{liu2014a,tian2015}. This assumption is held if the disk is axisymmetric. When the disk is non-axisymmetric, if \vR, \vPhi, and \vZ\ do not substantially change within the relative smaller range of the azimuth angle that the data samples expand, the invariance of the three mean velocity components is still approximately held. For our sample, the range of the azimuth angle is only about 20$^\circ$ and based on~\citet{kuijken1994} our Galaxy is not dramatically elliptical, then we can determine the three mean velocity components for each $R$ bin using the de-projection approach. 

\subsection{The velocity de-projection approach}\label{sect:methoddesc}

Although \citet{mcmillan2009} found that the de-projection can induce systematic bias in velocity ellipsoid, especially the cross-terms,~\citet{tian2015} have proved that the de-projection approach can well reproduce the first-order momenta, i.e. the mean velocity components, without taking into account the bias of the spatial sampling.

Given a star, the line-of-sight velocity can be obtained from 
\begin{eqnarray}\label{eq:deproject}
v_{los}=-\langle v_R\rangle\cos(l+\phi)\cos b \nonumber &+& \langle v_\phi \rangle(z)\sin(l+\phi)\cos b \nonumber\\
 &+& \langle v_z\rangle\sin b-v_{\odot,los},
\end{eqnarray}
where $l$, $b$, and $\phi$ are the Galactic longitude, latitude and the azimuth angle with respect to the Galactic center, respectively. Noted that \vPhi\ is a declining function of $|z|$. ~\citet{bond2010} suggested the empirical relation between \vPhi\ and $|z|$ as 
\begin{equation}\label{eq:vphiz}
\langle v_\phi \rangle(z) = \langle v_\phi \rangle(z=0) - 19.2 |z|^{1.25}.
\end{equation}
It is derived from the SDSS proper motion data beyond $z=1$\,kpc. However, our data is within  $z=1$\,kpc. Therefore, we have to  extrapolate the relation to our data. It is also worthy to note that \citet{smith2012} also provided a few data points of \vPhi\ between $|z|=500$ and 1000\, pc. We tested to use both relations in equation ~\ref{eq:deproject} to derive the 3D averaged velocity components. The difference between the two relations are very small. Therefore, we decided to adopt the relation from ~\citet{bond2010} in this work.

The solar motion about the Galactic center projected to the line of sight can be written as
\begin{equation}\label{eq:solar}
v_{\odot,los}=U_\odot\cos l\cos b+(v_0+V_\odot)\sin l\cos b+W_\odot\sin b,
\end{equation}
where $U_\odot$, $V_\odot$, and $W_\odot$ are the three components of the solar motion w.r.t the local standard of rest (LSR) and $v_0$ is the circular speed of the LSR. We adopt the solar motion from~\citet{tian2015} as $(U_\odot,V_\odot,W_\odot)=(9.58, 10.52, 7.01)$\,kms$^{-1}$, the distance of the Sun to the Galactic center is $R_0=8$\,kpc, and the circular speed of LSR is $v_0=238$\,kms$^{-1}$ ~\citep{schoenrich12}.

\subsection{Method validation with the mock RC data}\label{sect:methodvalid}

Before we apply the velocity de-projection method to the observed data, we run a Monte Carlo simulation to test its validation. We borrow the 3D spatial positions from the observed RC samples and re-assign 3D mock velocities for each star such that we prepare a mock dataset with exactly the same spatial sampling and known velocity. We then use the velocity de-projection approach to derive the mean velocity components at various radii from the mock data and compare them with the input values to test the validation of the method and to derive the uncertainties of the derived mean velocity components owe to the biased spatial sampling of the survey.

\begin{figure}[!t]
\centering
\includegraphics[width=0.5\textwidth, trim=0.cm 1.3cm 0cm 0.5cm, clip]{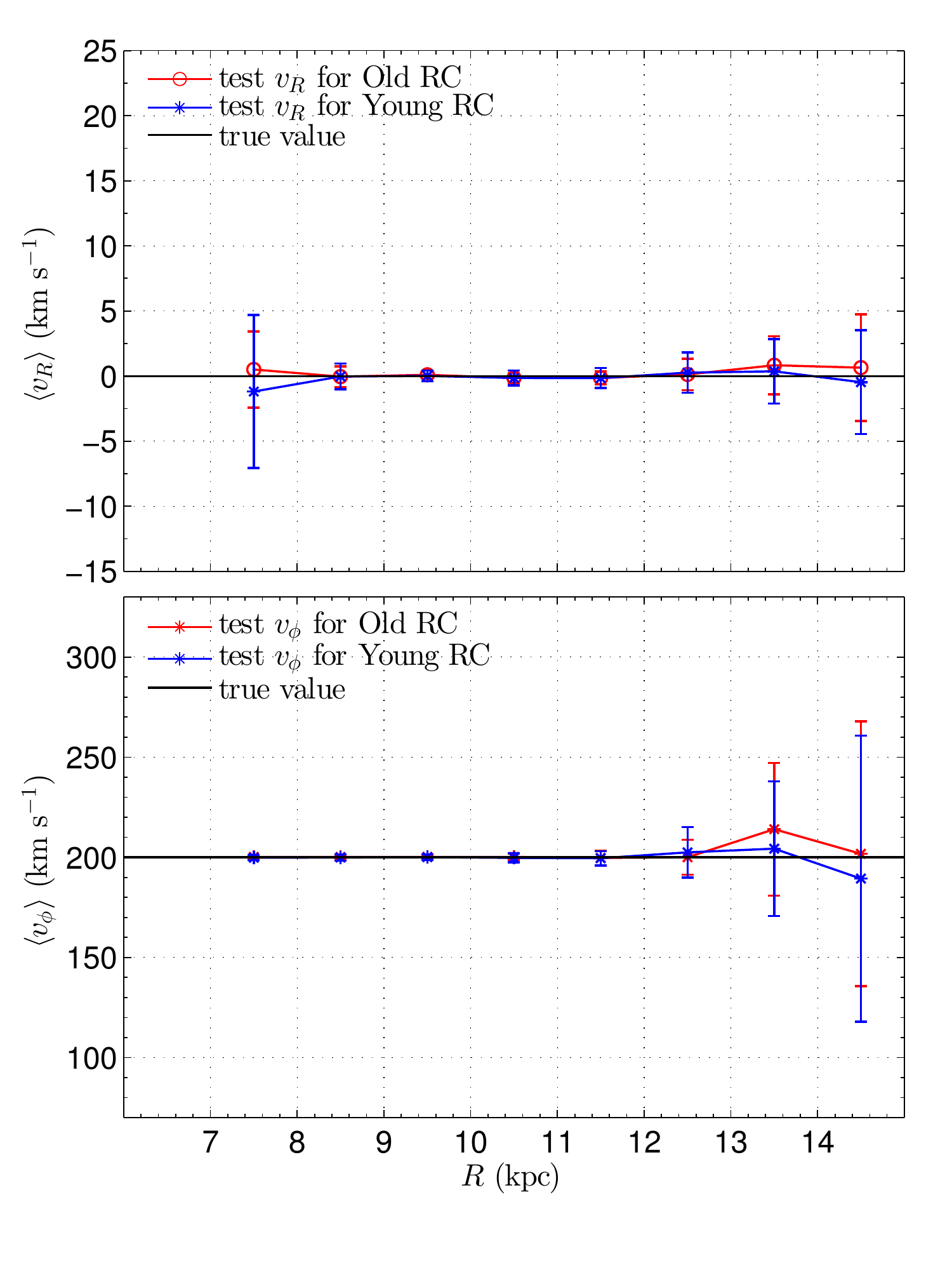}
\caption{The mean in-line velocities and their uncertainties derived from the velocity de-projection for the mock data. The top panel shows the variation of \vR\ with $R$ and the bottom shows the variation of \vPhi\ with $R$. The red and blue lines are the estimated velocities from old and young RC mock samples, respectively. The black lines indicate the input values when generating the mock data.}
   \label{fig:validate}
\end{figure}

For each mock star, we draw an arbitrary velocity vector from a 3D Gaussian distribution with the peak value of ($v_R$, $v_\phi$, $v_z$)=(0,200,0)\,\kms and sigmas of ($\sigma_R$, $\sigma_\phi$, $\sigma_z$)=(35,25,20)\,\kms. Then, for each $R$ bin we run a Markov chain Monte Carlo (MCMC) simulation to derive the most likely \vR, \vPhi, and \vZ\ and their uncertainties. We use the \emph{emcee} code to run the MCMC~\citep{forman2013}. We repeatedly produce 50 sets of the mock dataset with different random velocities drawn from the same Gaussian distribution and the final mean velocities as functions of $R$ for the young and old RC populations are shown in Figure~\ref{fig:validate}. The top panel shows the mean radial velocity as a function of $R$ for the young (blue asterisks) and old (red circles) mock RC stars. The de-projection-derived \vR\ for both populations well reproduce the true value of \vR, which is zero. The error bars at various $R$ bins are the standard deviations of \vR\ for the 50 simulations. The first point at $R=7.5$\,kpc and the points beyond 12\,kpc show larger errors, which is likely because the fewer samples falling in these bins. 

The bottom panel shows the similar test result for the azimuthal velocity. Again, the de-projection-derived \vPhi\ well follow the true value of 200\,kms. Compared with \vR, the error bars of \vPhi\ for the bins beyond 12\,kpc are significantly larger for two reasons: the fewer samples in these bins and the observed RC stars expand in a much smaller range in azimuth angle. Indeed, the range of azimuth angle for the RC stars located within $R=12$\,kpc is about 20-30$^\circ$ (see Figure~\ref{fig:density_distri}), while for the stars beyond 12\,kpc the range reduces to less than 10$^\circ$. The smaller range of $\phi$ is responsible for the larger uncertainties of \vPhi\ in these radii.

Nevertheless, this test demonstrates that the velocity de-projection approach is feasible to derive \vR\ and \vPhi\ with moderate uncertainties due to the sampling of the survey. It is worthy to note that the velocity de-projection also works out the mean vertical velocity with larger uncertainty. More investigations on the vertical velocity will be in Liu et al. in preparation.

\section{Results}\label{sect:result}

\subsection{The radial variation of the in-plane velocities}\label{sect:radvar}
Both the young and old RC stars are not evenly distributed in $R$, as shown in Figure~\ref{fig:density_distri}, therefore, we need to carefully choose the binning strategy along $R$. In principle, we want the number of stars in each $R$ bin to be large enough for better statistics. On the other hand, we want that the resulting velocities between the  neighboring bins can be comparable. Therefore, we separate the $R$ bins with equal number of 1500 for the young and old RC stars respectively. At the smallest and largest ends of $R$, namely when $R<8$\,kpc and when $R>12$\,kpc,  the number of stars become much fewer, keeping 1500 stars within a bin will significantly enlarge the bin size. Therefore, we reduce the number of stars in a bin at the two ends of $R$ to a much smaller number from 250 to 700. The choice of these numbers of stars at the two ends are empirical based on many tests so that the statistics and the resolution in $R$ in this two regimes can be relatively balance.

\citet{tian2015} have claimed that the LAMOST line-of-sight velocity is systematically smaller than the APOGEE measure by 5.7\,\kms\ for unclear reason. Thus, we add 5.7\,\kms\ to the line-of-sight velocity for all the RC stars before deriving the mean in-plane velocity components.

We then apply the velocity de-projection technique to the observed dataset and obtain the mean velocity components at each $R$ bin, as shown in the left panels of Figure~\ref{fig:mv_r}. In this work, we focus only on the in-plane velocities, i.e. \vR\ and \vPhi. The vertical velocity is beyond the scope of this paper and will be further discussed in the next paper (Liu et al. in preparation).

\subsubsection{Radial variation of \vR}\label{sect:varvr}
\begin{figure*}[!t]
\begin{center}
 \includegraphics[width=1.0\textwidth, trim=0.0cm 0.05cm 0.05cm 0.0cm, clip]{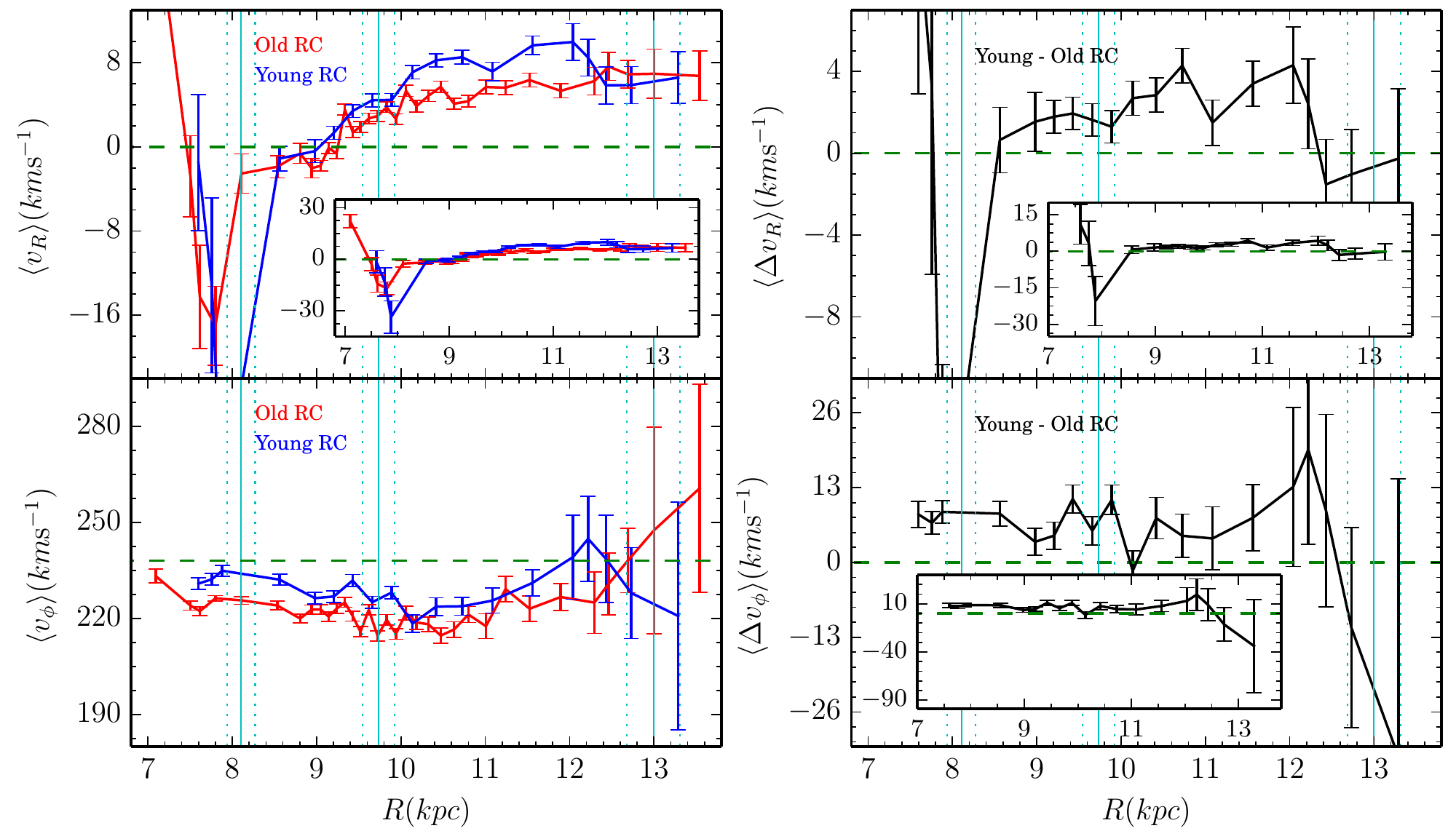}
 \caption{The top-left panel displays the variation of \vR\ for the young (blue) and old (red) RC stars. The green dashed values indicate the zero line. The bottom-left shows the variation of \vPhi\ for the both populations. The green dashed line indicates the 238\,\kms\ as a reference. The top-right panel shows the differential radial velocity (Young - Old) along $R$. And the bottom-right panel shows the differential azimuthal velocities as a function of $R$. The cyan vertical solid and dash lines in all panels mark the range of the location of spiral arms~\citep{reid2014}. The corresponding values in all four panels are specified in Table \ref{tab:mv_r1} and \ref{tab:mv_r2}.
 }
   \label{fig:mv_r}
\end{center}
\end{figure*}

The top-left panel of Figure~\ref{fig:mv_r} shows the radial variation of \vR\ for both the young (blue lines) and the old (red lines) RC stars. The error bars are composed of two parts: the uncertainty of the derived \vR\ from the MCMC procedure and the uncertainties of the sampling in the $R$ bin obtained from the Monte Carlo simulation mentioned in section~\ref{sect:methodvalid}. The detail values showing in the figure can be found in Table~\ref{tab:mv_r1} for the young and Table~\ref{tab:mv_r2} for the old RC stars.

The most prominent feature displayed in the panel is that \vR\ for the both populations are mostly negative at $R\lesssim9.0$\,kpc (except only one point at $R=7.09$\,kpc, which is probably a real feature but for some unclear reasons, or just due to the special selection effect at this radius), while both the radial profiles of \vR\ become positive beyond $R\sim9.0$\,kpc. For the bins with $R\lesssim9.0$\,kpc, the mean radial velocity can be as large as $\sim-33$\,\kms\ at $R\sim7.6$\,kpc. For the bins with $R\gtrsim9.0$\,kpc, most of the mean radial velocities are between 4 and 8\,\kms. It is obvious that the young and the old RC stars have different radial \vR\ profile, especially at $9<R<12$\,kpc. Although the two populations show different \vR\ at different $R$ bins, the cross-zero points of the radial profiles of \vR\ are consistent with each other. 

Compare this result with \citet{bovy2015}, who mapped the line-of-sight velocity into $X$-$Y$ plane (see their Fig.2), the pattern of the radial variation is quite consistent with this work. If we only focus in about 1\,kpc around the Sun, the variation of the radial velocity with $R$ is also in agreement with~\citet{siebert2011a}. At the location of the Sun, their $v_R$ reads about  $-7\sim-24$\,\kms, while we derive about $-5\sim-33$\,\kms. We then look at the variation of \vR\ with~\citet{carlin2013} and find that within $R\sim9$\,kpc, the radial velocities is about $-10$\,\kms\ and at $R>9$\,kpc, the radial velocity increases to about $-5$\,\kms. The trend of $v_R$ with $R$ in their result seems similar to this work, while the values of the velocity shift by about $-10$\,\kms\ contrast to our RC stars. Notice that~\citet{carlin2013} measured the radial velocity from the combination of the line-of-sight velocities and the proper motions, complicated systematics may be responsible for the difference. For instance, \citet{carlin2013} did not correct the LAMOST DR1 derived line-of-sight velocities by adding 5.7\,\kms, which is found by comparing with the APOGEE data~\citep{tian2015}; the different distance estimations may induce systematics; and the proper motion may also lead to some bias in velocity.

The differential radial velocity, \dvR,  between the young and the old populations are displayed in the top-right panel of Figure~\ref{fig:mv_r} and the detailed values are listed at the last few columns in Table~\ref{tab:mv_r1}. The plot shows that the mean radial velocities for the young RC stars are larger than those for the old RC stars by about  3\,\kms\ between $R=9$ and 12\,kpc. This difference is at least 2-$\sigma$ significance for each bin. However, combined with the all the bins within the range of $R$, the difference turns to be very significant, implying that the difference should be a real feature. It is worthy to note that the range of $R$ happened to be located between the Perseus and the Outer arm. The radial positions of the spiral arms, i.e. the local arm, the Perseus arm, and the Outer arm, from the trigonometric parallaxes of the masers~\citep{reid2014} are superposed in the figure with cyan vertical lines.

Another large difference between the two populations of RC stars occurs at $R\sim 8$\,kpc, at which \vR\ for the young RC stars is significantly smaller than that of the old stars and hence produces a huge dip.

\subsubsection{Radial variation of \vPhi}\label{sect:varvphi}
The bottom-left panel of Figure~\ref{fig:mv_r} displays the radial profiles of \vPhi\ for the young (blue line) and old (red line) RC stars. The dashed horizontal line indicates the velocity of 238\,\kms, which is the adopted circular speed in the solar neighborhood, as a reference.

The radial profile of \vPhi\ shows two patterns. In the regime of $R<11$\,kpc, the mean azimuthal velocities for both the young and old populations mildly decrease by 10-20\,\kms\ from $R\sim7.0$\,kpc to 10.5\,kpc. Moreover,  \vPhi\ for the young population is larger than those for the old population by $\sim10$\,\kms\ in this regime. This is natural since the young RC stars should be kinematically colder and hence rotates faster than the old population. However, beyond $R=10.5$\,kpc, the mean azimuthal velocities for the old RC stars increase to around 260\,\kms\ at $R>13$\,kpc, while the young population abruptly increases to 245\,\kms\ at $R=12$\,kpc, and then drop back to 220\,\kms\ when $R>13$\,kpc. The abrupt change of \vPhi\ for the young population is within 1$\sigma$ uncertainty, thus must be due to the fewer observed sample in such large distance. The pattern of the radial profile of \vPhi\ for the old population is consistent with the the result from~\citet{liu2014a}, who used mostly the old population of the RC stars to reconstruct the in-plane kinematics of the Galactic outer disk. The reason of the sudden raise of \vPhi\ for old population is not quite clear. 

The bottom-right panel of Figure~\ref{fig:mv_r} shows the radial profile of the differential mean azimuthal velocities, \dvPhi, subtracting \vPhi\ for the old stars from that of the young stars. \dvPhi\ is essentially flat at about 10\,\kms, reflecting the difference in rotation between the two age populations.


\section{Discussions}\label{sect:disc}
The radial variation of \vR\ discussed in section~\ref{sect:varvr} looks much clearer than that of \vPhi\ shown in section~\ref{sect:varvphi}, because the uncertainty in the derivation of \vR\ is much smaller for our dataset. From \vR\ profiles, we identify two types of peculiar velocity. First, both the young and old populations have negative \vR\ with $R\lesssim9.0$\,kpc and turn to be positive beyond the value. If we consider this type of variation of the radial velocity as a radial wave, the wave length must be larger than 6\,kpc by a factor of a few, since there is only one cross-zero point between $R=8$ and 14\,kpc. In this sense, we denote such a peculiar velocity as a \emph{long-wave} mode. Second, the differential radial velocity shows a bump between $R=9$ and 12\,kpc, which is much shorter than the \emph{long-wave} mode. We then denote the wave shown in the difference of \vR\ between the young and old populations as a \emph{short-wave} mode. In the next two sub-sections we discuss the possible mechanisms for the two types of the peculiar velocities.

\subsection{The \emph{long-wave} mode}\label{sect:longwave}

From previous works, it is noted that the velocity wave perturbed by the spiral structures are always spatially correlated with the spiral arms \citep[see][]{siebert2012,faure2014,grand2015}. The \emph{long-wave} mode structure in radial velocity, however, does not show correlation with any spiral arm in the outer disk. Moreover, it seems that the perturbation from the merging dwarf galaxies, e.g. the Sgr dwarf, will raise vertical wave but may not intensively affect the in-plane velocity~\citep{gmo+13}. Therefore, the \emph{long-wave} mode may neither be induced by the spiral arms nor by the merging dwarf galaxies.

Alternatively, if the disk is elliptical due to the bar, the orbital ellipticity can naturally produce non-zero mean radial velocities. It is apparently that the elliptical gravitational potential introduced by \citet{kuijken1994} and \citet{rix1995} do not match the radial variation of \vR. Then the rotating bar, which makes the elliptical gravitational potential rotate, is taken into account. 

\begin{figure}[!t]
\begin{center}
 \includegraphics[width=0.5\textwidth, trim=0.5cm 2.2cm 1.0cm 1.5cm, clip]{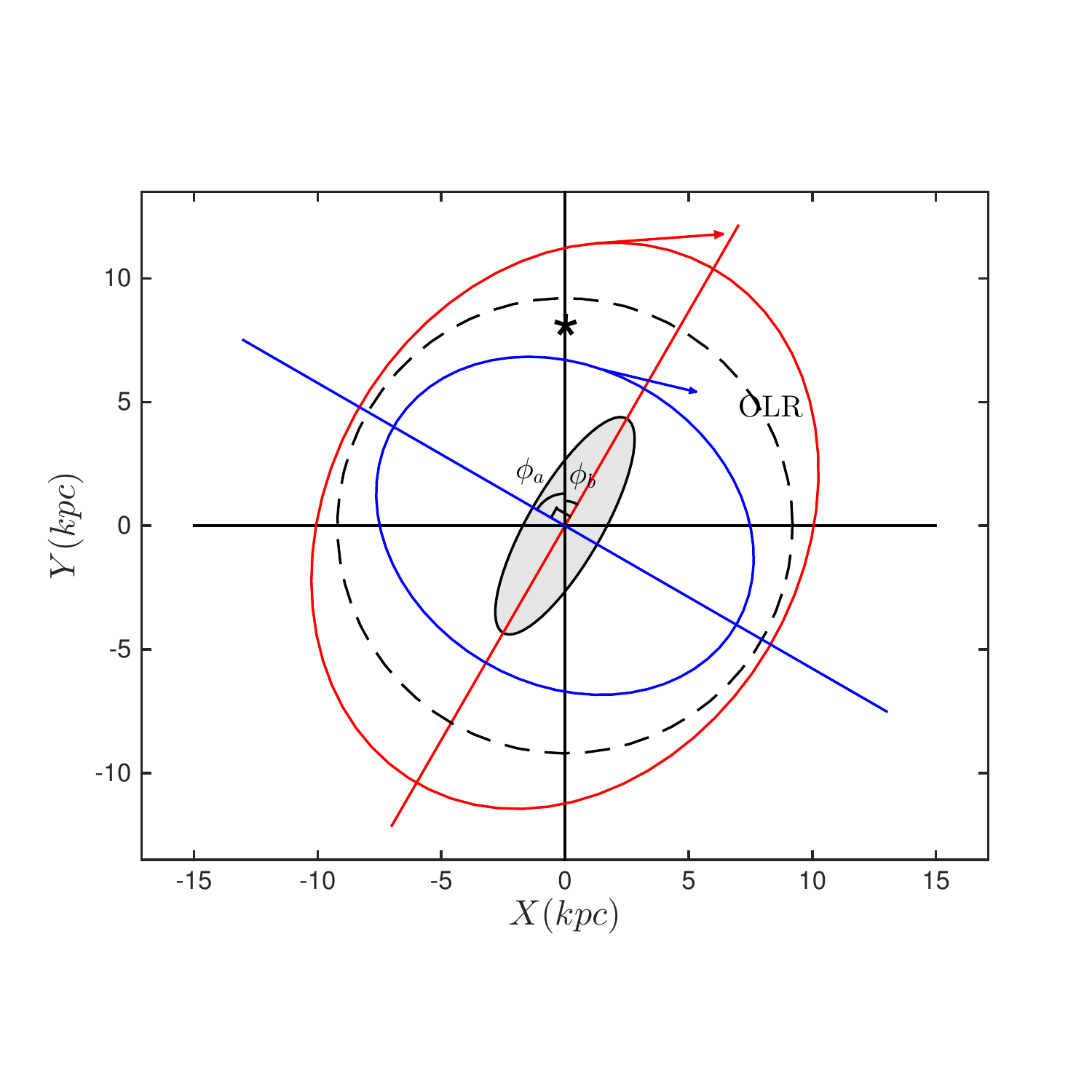}
 \caption{This cartoon demonstrates the elliptical closed loop orbit inside (blue curve) and outside (red curve) the OLR. The red and blue straight lines represent for the major axes of the two orbits, respectively. The arrows indicate the velocity vectors at the Galactic anti-center direction. $\phi_b$ is the angle of the bar from the Galactic center-Sun baseline, while $\phi_a$ is angle of the minor axis of the bar. The location of OLR and the Sun are marked as the black dashed circle and asterisk, respectively.
 }
   \label{fig:ellip}
\end{center}
\end{figure}

Notice that the direction of the major axis of the elliptical closed loop orbits in a rotating elliptical gravitational potential changes a few times at various radii (see section~\ref{sect:intro}) To better demonstrate this we show a cartoon in Figure~\ref{fig:ellip}. The elliptical closed loop orbits  (shown as the blue line) located between the CR and OLR should be perpendicular to the major axis of the bar. In this case, the radial component of the velocity (indicated with blue arrow) for the closed loop orbit at the Galactic anti-center region should be negative according to the relative spatial position between the bar and the Sun. On the other hand, the orbits located beyond the OLR (red circle) are aligned with the major axis of the bar and thus show positive radial velocity (indicated with the red arrow) in the Galactic anti-center direction. Right at the OLR, the two directions of the elliptical closed orbits mixed with each other and together show zero mean radial velocity. Therefore, it is expected that the closed loop orbit should be elliptical at both sides of the OLR, but keep circular right at the OLR. Meanwhile, looking the stars along the Galactic anti-center direction from the Sun, the radial velocity is expected to be negative at smaller $R$, and be zero at this point corresponds to the OLR, and then becomes positive at larger $R$.

For the real stars, they are rarely moving exactly on the closed stellar orbits, but on the non-closed objects with radial excursions. The excursions in radial direction may lead to dispersions and may produce an envelope surrounding the closed loop orbit. However, the mean velocities should well follow that on the  closed loop orbits.

This schematic scenario with a rotating bar in the central region can well explain the observed radial variation of the mean radial velocities shown in Figure~\ref{fig:mv_r}.  

\begin{figure}[!t]
\begin{center}
 \includegraphics[width=0.5\textwidth, trim=0.0cm 0.03cm 0.03cm 0.0cm, clip]{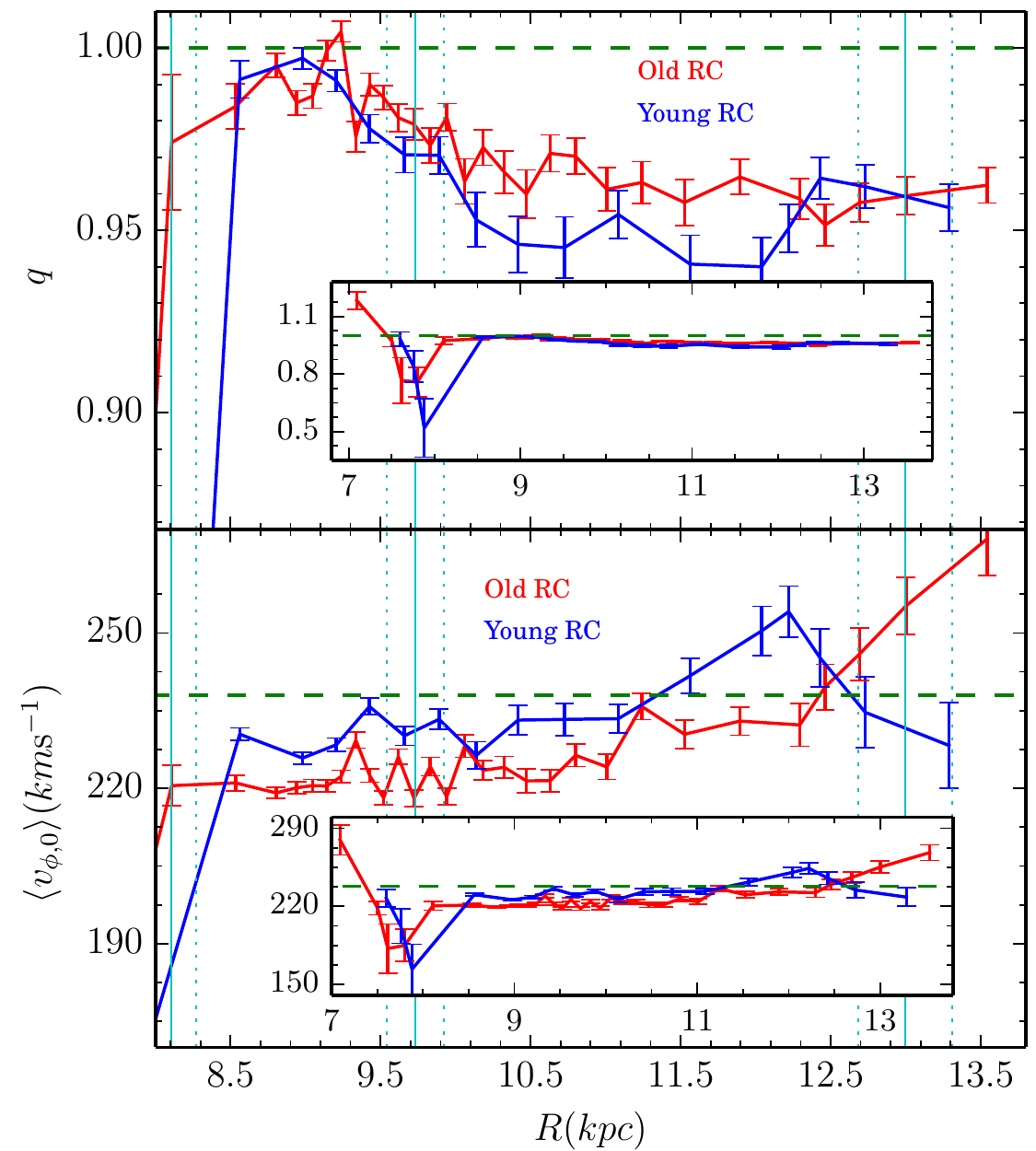}
 \caption{The top panel displays the derived axis ratio from the young (blue) and old (red) RC stars. The bottom panel shows the axisymmetric azimuthal velocity derived from the young (blue) and old (red) RC stars. These quantities are derived given $\phi_b = 20 \degr$. The cyan and green lines are same as in Figure~\ref{fig:mv_r}.}
   \label{fig:q_vc_r}
\end{center}
\end{figure}

We then estimate the variation of the ellipticity due to the rotating bar with the simple projection of the in-plane velocity components for the stars with elliptical closed loop orbits, which should exist in the underlying gravitational potential with a rotating bar:
\begin{eqnarray}\label{eq:mean_v}
\langle v_R\rangle &=&  \epsilon v_{\phi,0} \langle \sin2(\phi - \phi_0)\rangle\nonumber\\
\langle v_{\phi}\rangle &=& v_{\phi,0}(1 - \epsilon \langle \cos 2(\phi - \phi_0)\rangle),
\end{eqnarray}
where \VPhi\ is the azimuthal velocity when the closed loop orbit is circular. $\epsilon$ refers to the flattening of the elliptical closed loop orbit. And $\phi_0$ is the angle between the major axis of the bar and the line connecting the Galactic center and the Sun. The axis ratio of the orbit can be written as
\begin{equation}\label{eq:q}
q = 1 - \epsilon.
 \end{equation}
The axis ratio $q$ and \VPhi\ can be derived from the observed \vR\ and \vPhi\ when $\phi_0$ is known. We give three measurments for $\phi_0=20^\circ$, $30^\circ$, and $40^\circ$, respectively. The results corresponding to various $\phi_0$ are quite similar and are listed in Tables~\ref{tab:mv_r1} and~\ref{tab:mv_r2}. Figure~\ref{fig:q_vc_r} shows the derived $q$ and \VPhi\ at various $R$ bins for both the young (blue lines) and old (red lines) RC stars only at $\phi_0=20^\circ$. Focusing only on the old population since it is less affected by other types of perturbation, the axis ratio of the closed loop orbits shown in the top panel is about 0.8 with $R<8$\,kpc and then increases to almost 1.0, i.e. the closed loop orbits are circular, at the cross-zero point ($R = 9.24\pm0.2$\,kpc), and then beyond 9.24\,kpc, the axis ratio drops again to around 0.95. The difference between the old and young populations should be related to the \emph{short-wave} mode, which will be discussed in the next sub-section. The radial variation of the axisymmetric azimuthal velocity \VPhi\ is shown in the bottom panel. Both the young and old populations show quite flat \VPhi\ with $R<10.5$\,kpc. Beyond this radius, due to the unclear abrupt changes, the derived \VPhi\ shifts far from the reference 238\,\kms. 

The derived $q$ of the old population is evident that the closed loop orbits are elliptical before and after $R = 9.24\pm0.2$\,kpc, but are circular at this radius. 

For the rotating bar, the CR should be in the inner disk of $R\sim4-6$\,kpc depending on the pattern speed of the bar~\citep{gerhard2011}. Then the OLR should be beyond $R_0$ if the circular speed is around 238\,\kms. Therefore, the point $R = 9.24\pm0.2$\,kpc with $q\sim1.0$ and \vR$=0$ should be exactly at the OLR. Consequently, the circular  frequency, $\Omega$, at this point should satisfy the following equation:
\begin{equation}\label{eq:olr}
m(\Omega-\Omega_b)=\kappa,
\end{equation}
where $m=2$ in our case, $\kappa$ is the radial frequency of the stellar orbit at OLR, and $\Omega_{b}$ is the pattern speed of the bar~\citep[cf. Eq. 3.150 in ][]{binney2008}. With the assumption that the epicyclic approximation is valid for the stellar orbits, $\kappa$ can be given as
\begin{equation}\label{eq:kappa}
\kappa^2=(R{d\Omega^2\over{dR}}+4\Omega^2).
\end{equation}
According to \citet{binney2008}, it can be derived that $\kappa=\sqrt{3}\Omega$ given the flat circular speed nearby the OLR. With known local circular speed (238\,\kms), the pattern speeds of the bar is estimated as 48$\pm3$\,\kms\ikpc, within the range of the previous estimates~\citep{gerhard2011,long2013,wang2013}.

The \emph{long-wave} mode is one of the most prominent features compared with the \emph{short-wave} mode discussed in next section. And it may reflect the ellipticity of the disk induced by the rotating bar according to our above discussions. This can well explain that when \citet{grand2015} compared the power spectra from different simulations with that from the APOGEE RC data derived by~\citet{bovy2015}, they found that the simulations with a bar show very similar power spectra to the observed one. 

It is noted that the Eq. (3.150) in~\citet{binney2008} is based on the epicyclic approximation and hence requires that the velocity dispersion must be sufficiently small. Indeed, the bottom-left panel of Figure~\ref{fig:mv_r} shows that the azimuthal velocities for both RC populations are quite close to the circular speed with lags of 10-20\,\kms, implying that both the populations are kinematically cold and the epicyclic approximation should be held. Therefore, we would  expect that the ellipticity of the two RC populations should be same.

\subsection{The \emph{short-wave} mode}\label{sect:shortwave}

Another interesting feature shown in Figure~\ref{fig:mv_r} is that the mean radial velocities between the young and the old RC stars are significantly different between 9 and 12\,kpc. Because the Perseus arm is located at about 10\,kpc and the Outer arm is located at about 13\,kpc in the Galactic anti-center region~\citep{reid2014}, the differential radial velocity may be associated with the two spiral arms.

In order to investigate whether the differential velocity is associated with the spiral structures, we adopt the analytical analysis discussed by~\citet{siebert2012}, which is based on the perturbation theory of the density wave. The potential from the spiral structures can be given in the following form:
\begin{equation}\label{eq:spiralpot}
\Phi_{spiral}=\mathcal{A}(R)\exp[i(\omega t-m\phi+\Theta(R))],
\end{equation}
where $\mathcal{A}(R)$ is the amplitude of the perturbation, $m$ is the number of arms, $\omega$ is the angular speed, which is associated with the angular frequency via  $\Omega_s=\omega/m$, $\phi$ is the azimuth angle, and $\Theta(R)$ is a monotonic function to describe the shape of the spiral. The spiral structures can induce the perturbed radial velocity ($\langle\widetilde V_R\rangle$) and azimuthal velocity ($\langle\widetilde V_\phi\rangle$) as
\begin{eqnarray}\label{eq:spiralpert}
\langle \widetilde V_R\rangle &=& \frac{k\mathcal{A}}{\kappa}\frac{\nu}{1-\nu^2}\mathcal{F}_{\nu}^{(1)}(x)\cos(\chi);\nonumber\\
\langle \widetilde V_{\phi}\rangle &=& - \frac{k\mathcal{A}}{2\Omega}\frac{1}{1-\nu^2}\mathcal{F}_{\nu}^{(2)}(x)\sin(\chi).
\end{eqnarray}
where $x = {k^2\sigma_{R}^2}/{\kappa^2}$, $k$ is the radial wavenumber, $\sigma_R$ is the radial velocity dispersion, $\nu$ is defined by $m(\Omega_p - \Omega)/\kappa$, the radial frequency $\kappa = -4B\Omega$ and the circular frequency $\Omega=A - B$. Here, $A$ and $B$ are the Oort's constants.

$\mathcal{F}_{\nu}^{(1)}$ and $\mathcal{F}_{\nu}^{(2)}$ are the reduction factors that reflect how sensitivity the stars responding to the perturbation. They are related to the velocity dispersions. In the extreme case in which velocity dispersion is zero, $\mathcal{F}_{\nu}^{(1)}=\mathcal{F}_{\nu}^{(2)}=1$, while in the opposite extreme case in which the dispersion is so large that the stars do not respond to the perturbation at all,   $\mathcal{F}_{\nu}^{(1)}=\mathcal{F}_{\nu}^{(2)} \to 0$. $\chi$ is defined by 
\begin{equation}\label{eq:chi1}
\chi = \omega t - m\phi + \Theta(R).
 \end{equation} 
In the case of logarithmic spirals, $\Theta(R) = m \cot i \ln R$, where $i$ is the pitch angle. Then Eq. (\ref{eq:chi1}) can be written in more convenient form:
 \begin{equation}\label{eq:chi2}
\chi = \chi_0 + m(\cot i \ln(R/R_0) - (\phi - \phi_0)).
 \end{equation} 
 The quantities with subscript 0 in the equation refer to the value at the Sun's location. The radial wavenumber $k$ is then given by 
  \begin{equation}\label{eq:chi2}
k(R) = {d\Theta(R)\over{dR}} = m\cot i /R,
 \end{equation} 
with $k(R)<0$ for trailing waves and $k(R)>0$ for leading waves.

\begin{figure}[!t]
\begin{center}
 \includegraphics[width=0.7\textwidth, trim=0.5cm 0.5cm 0.5cm 0.5cm, clip]{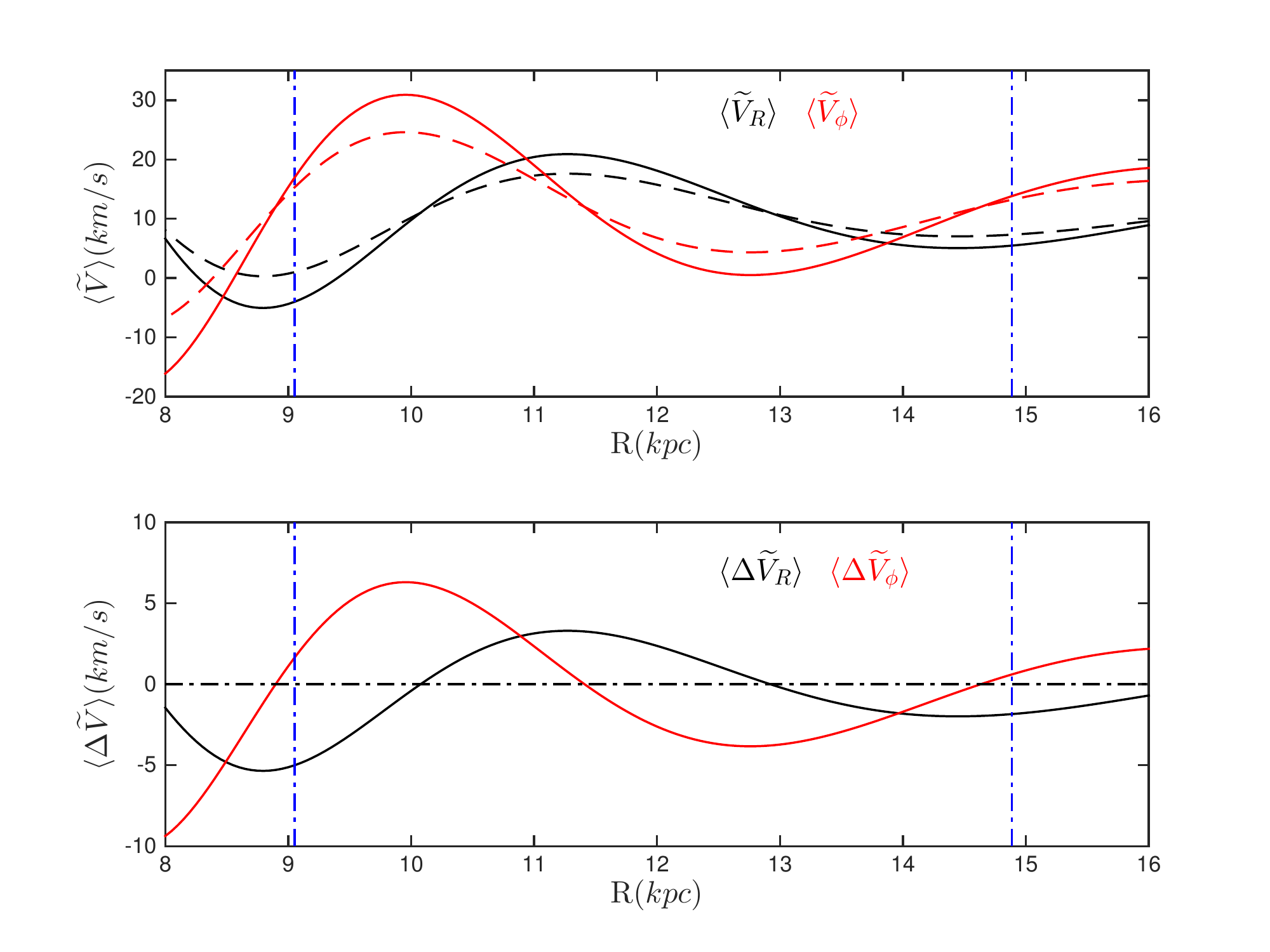}
 \caption{The top panel displays the peculiar velocities derived from the toy model of the perturbation of the density wave spiral arms. The solid curves are for the mimic-young population and the dashed curves are for the mimic-old population. The black lines stand for the mean radial velocity $\langle\widetilde V_R\rangle$,  and the red lines are for $\langle\widetilde V_\phi\rangle$. The bottom panel shows the differential mean radial velocity (black) and mean azimuthal velocity (red) between the mimic-young and old populations as functions of $R$. The blue dot-dashed lines indicate the location of the spiral arms in the model.
 }
   \label{fig:Vosci}
\end{center}
\end{figure}

We generate the $\langle\widetilde V_R\rangle$ and $\langle\widetilde V_\phi\rangle$ for a toy model of the Galactic potential including an exponential disk and 2 spiral arms based on Eq. (\ref{eq:spiralpot}) for comparison with the observed \emph{short-wave} mode. The adopted parameters of the model of potential are listed in Table~\ref{tab:spiralmodel}. The top panel of Figure~\ref{fig:Vosci} shows the radial variations of the model $\langle\widetilde V_R\rangle$ (black lines)  and $\langle\widetilde V_\phi\rangle$ (red lines) in the Galactic anti-center direction. The solid lines represent for the radial variations mimic to the young population with $\mathcal{F}^{(1)}=\mathcal{F}^{(2)}=0.8$, while the dashed lines stand for the radial variations mimic to the old population with $\mathcal{F}^{(1)}=\mathcal{F}^{(2)}=0.5$. The dot-dashed blue lines refer to the two spiral arms in the toy model. It shows that the phases of the radial and azimuthal velocity, relative to the location of the spiral arms, are quite different. For the radial velocity, the peak of the wave occurs in between the two arms and the valleys of the wave well overlap the location of the spiral arms, while for the azimuthal velocity, the peak of the wave is not consistent with the radial velocity, but right behind of the spiral arm. And as expected, the young population has larger amplitude than the older population in both velocities.

The bottom panel of Figure~\ref{fig:Vosci} shows the differential velocities between the young and the old populations. Compared this with the differential velocities in the right panels of Figure~\ref{fig:mv_r}, we find that, qualitatively, they show quite similar pattern in the differential radial velocity to the observed RC stars. Therefore, it is very likely that the significant differential radial velocity for the young and old RC stars reflect the different extents of response of the perturbation of the two outskirt spiral arms. However, for the observed differential azimuthal velocity, it is not agree with the toy model. This is probably because that, in the realistic situation, the perturbation in the velocities are much more complicated and the measured velocities may be the combination of all perturbations.

\section{Conclusions}\label{sect:conclusions}

The perturbations due to the bar, the spiral arms, the merging dwarf galaxies etc. may only give very small amount of offset in the stellar kinematics. The disk stars are likely affected simultaneously by a few different kinds of perturbations. Hence, the resulting variations of the mean velocity may be the combination of different modes. In this work, we identify at least two types of the perturbation in the LAMOST RC stars. The \emph{long-wave} mode seems associated with the rotating bar, which reshapes the disk to be elliptical. In the underlying rotating elliptical gravitational potential, the closed loop orbits corresponding to the minimum total energy is no longer circular as in an axisymmetric potential, but become elliptical. Moreover, the major-axis of the elliptical orbits varies with radii due to the torque induced by the rotating non-axisymmetric potential. This leads to a various radial velocity along the Galactic anti-center direction depending on the relevant angular position between the Sun the the bar. The radial velocity is negative within the OLR and become positive beyond the OLR. At the OLR, the radial velocity is exactly zero. The observed radial profile of \vR\ perfectly shows such a trend. The derived axis ratio of the elliptical orbits is about 0.8 within OLR and 0.95 beyond for the case of $20^\circ$. Moreover, the \vR$=0$ point, which should be at the OLR, implies that the pattern speed of the bar is $48\pm3$\,\kms\ikpc\ with the circular speed of 238\,\kms.

The differential radial velocity between the young and old RC stars may reveal the perturbation from the spiral arms. The amplitude of the perturbation induced by the spiral structures depends on how cold the kinematics of the stars is, i.e. the stars with smaller velocity dispersion are easier to be perturbed by the spiral arms, while the torque from the rotating bar affects all ages of the stars with large angular momenta. Therefore, \dvR\ between the young and old populations can \emph{filter} the effect from the bar and emphasize the influence from the spiral arms. Based on the simplified toy model of the perturbation from density wave spiral structures, we find the differential radial velocity is very likely associated with the spiral arms, although the differential azimuthal velocity does not completely coincide with the perturbation model.

\acknowledgements
This work is supported by the Strategic Priority Research Program ``The Emergence of Cosmological Structures" of the Chinese Academy of Sciences, Grant No. XDB09000000 and the National Key Basic Research Program of China 2014CB845700. H-J.T acknowledges the National Natural Science Foundation of China (NSFC) under grants 11503012 and U1331202. C. L. acknowledges the NSFC under grants 11373032 and 11333003, Y. W. acknowledges NSFC under grant 11403056. Guoshoujing Telescope (the Large Sky Area Multi-Object Fiber Spectroscopic Telescope LAMOST) is a National Major Scientific Project built by the Chinese Academy of Sciences. Funding for the project has been provided by the National Development and Reform Commission. LAMOST is operated and managed by the National Astronomical Observatories, Chinese Academy of Sciences.

\bibliographystyle{apj}
\clearpage
\begin{turnpage}
\begin{table*}
\caption{The averaged velocities in each $R$ bin for the young RC stars and the differential velocities between the young and old populations, corresponding to the Figure~\ref{fig:mv_r}.}\label{tab:mv_r1}
\centering
\scriptsize
\begin{tabular}{c|c|c|c|c|c|c|c|c|c|c|c}
\hline\hline
$ N $& $\langle R \rangle$ & \vR & \vPhi & \multicolumn{3}{c|}{\VPhi} & \multicolumn{3}{c|}{$q$}&$\Delta\langle v_R\rangle$&$\Delta\langle v_{\phi}\rangle$\\
\hline
 &kpc &\multicolumn{2}{c|}{\kms} &\multicolumn{3}{c|}{\kms}& \multicolumn{3}{c|}{} & \multicolumn{2}{c}{\kms}\\
 \hline
 \multicolumn{4}{c|}{Young RC}& \multicolumn{3}{c|}{Young RC}& \multicolumn{3}{c|}{Young RC} & \multicolumn{2}{c}{Young - Old}\\
\hline
\multicolumn{4}{c|}{}& $\phi_0=20\degr$&$\phi_0=30\degr$&$\phi_0=40\degr$& $\phi_0=20\degr$&$\phi_0=30\degr$&$\phi_0=40\degr$& \multicolumn{2}{c}{}\\
\hline
250&7.60&-1.49$\pm$8.64&230.87$\pm$2.23&227.19$\pm$7.77&228.97$\pm$4.11&229.94$\pm$2.27&0.98$\pm$0.04&0.99$\pm$0.02&0.99$\pm$0.02&11.05$\pm$10.33&8.33$\pm$2.74\\
250&7.76&-13.18$\pm$10.08&232.22$\pm$2.21&202.22$\pm$15.48&217.58$\pm$7.26&225.21$\pm$3.73&0.84$\pm$0.08&0.91$\pm$0.04&0.93$\pm$0.03&3.19$\pm$11.11&6.87$\pm$2.50\\
250&7.88&-33.70$\pm$11.00&234.89$\pm$2.18&163.61$\pm$22.31&201.03$\pm$9.55&219.05$\pm$5.08&0.52$\pm$0.14&0.76$\pm$0.06&0.83$\pm$0.04&-20.44$\pm$11.95&8.78$\pm$2.48\\
1500&8.56&-1.11$\pm$1.57&232.11$\pm$2.26&230.45$\pm$1.23&231.27$\pm$0.86&231.77$\pm$0.73&0.99$\pm$0.01&0.99$\pm$0.00&0.99$\pm$0.00&0.65$\pm$2.08&8.48$\pm$2.86\\
1500&8.98&-0.40$\pm$1.50&226.34$\pm$2.39&225.84$\pm$1.15&226.10$\pm$1.06&226.26$\pm$1.04&1.00$\pm$0.00&1.00$\pm$0.00&1.00$\pm$0.00&1.54$\pm$1.95&3.57$\pm$3.00\\
1500&9.21&1.34$\pm$0.77&226.86$\pm$2.28&228.41$\pm$1.31&227.61$\pm$1.21&227.08$\pm$1.18&0.99$\pm$0.00&0.99$\pm$0.00&0.99$\pm$0.00&1.79$\pm$0.97&4.64$\pm$2.93\\
1500&9.43&3.40$\pm$0.76&231.85$\pm$2.33&235.80$\pm$1.60&233.77$\pm$1.37&232.41$\pm$1.32&0.98$\pm$0.00&0.98$\pm$0.00&0.99$\pm$0.00&1.95$\pm$0.97&11.05$\pm$2.98\\
1500&9.66&4.43$\pm$0.76&225.07$\pm$2.31&230.15$\pm$1.84&227.52$\pm$1.45&225.77$\pm$1.36&0.97$\pm$0.00&0.98$\pm$0.00&0.98$\pm$0.00&1.63$\pm$0.97&5.50$\pm$2.99\\
1500&9.89&4.46$\pm$0.76&228.10$\pm$2.38&233.33$\pm$1.97&230.63$\pm$1.57&228.86$\pm$1.48&0.97$\pm$0.01&0.98$\pm$0.00&0.98$\pm$0.00&1.30$\pm$0.97&10.88$\pm$3.10\\
1500&10.14&7.10$\pm$0.80&218.39$\pm$3.51&226.36$\pm$2.65&222.22$\pm$1.92&219.43$\pm$1.75&0.95$\pm$0.01&0.96$\pm$0.00&0.97$\pm$0.00&2.69$\pm$1.04&-1.47$\pm$4.53\\
1500&10.42&8.22$\pm$0.81&223.69$\pm$3.57&233.18$\pm$2.82&228.26$\pm$2.06&225.01$\pm$1.89&0.95$\pm$0.01&0.96$\pm$0.00&0.96$\pm$0.00&2.84$\pm$1.03&7.71$\pm$4.62\\
1500&10.73&8.52$\pm$0.81&223.80$\pm$3.62&233.31$\pm$3.10&228.38$\pm$2.23&225.04$\pm$2.04&0.95$\pm$0.01&0.96$\pm$0.00&0.96$\pm$0.00&4.28$\pm$1.04&4.63$\pm$4.74\\
1500&11.09&7.16$\pm$1.18&225.56$\pm$5.19&233.45$\pm$2.71&229.35$\pm$2.19&226.55$\pm$2.07&0.95$\pm$0.01&0.96$\pm$0.00&0.97$\pm$0.00&1.50$\pm$1.45&4.16$\pm$7.32\\
1500&11.56&9.65$\pm$1.18&231.13$\pm$5.35&241.72$\pm$3.40&236.21$\pm$2.71&232.45$\pm$2.55&0.94$\pm$0.01&0.95$\pm$0.00&0.96$\pm$0.00&3.40$\pm$1.45&7.75$\pm$7.53\\
557&12.04&9.97$\pm$2.36&239.16$\pm$18.16&250.40$\pm$4.75&244.56$\pm$4.32&240.64$\pm$4.23&0.94$\pm$0.01&0.95$\pm$0.00&0.96$\pm$0.00&4.31$\pm$2.50&13.10$\pm$18.95\\
557&12.22&8.49$\pm$2.36&244.86$\pm$18.28&254.11$\pm$4.96&249.27$\pm$4.72&245.97$\pm$4.68&0.95$\pm$0.01&0.96$\pm$0.00&0.97$\pm$0.00&2.41$\pm$2.95&19.59$\pm$22.43\\
557&12.43&5.85$\pm$2.37&238.64$\pm$18.51&245.15$\pm$5.60&241.76$\pm$5.49&239.47$\pm$5.47&0.96$\pm$0.01&0.97$\pm$0.00&0.98$\pm$0.00&-1.52$\pm$2.95&8.98$\pm$22.66\\
557&12.73&5.87$\pm$2.37&227.95$\pm$18.93&234.67$\pm$6.85&231.19$\pm$6.77&228.87$\pm$6.75&0.96$\pm$0.01&0.97$\pm$0.00&0.97$\pm$0.00&-1.03$\pm$2.95&-11.33$\pm$23.14\\
557&13.29&6.58$\pm$3.40&220.76$\pm$49.73&228.28$\pm$8.30&224.39$\pm$8.21&221.79$\pm$8.20&0.96$\pm$0.01&0.97$\pm$0.00&0.97$\pm$0.00&-0.26$\pm$4.71&-33.74$\pm$67.43\\

\hline
\hline
\end{tabular}
\end{table*}
\end{turnpage}
\clearpage

\begin{table*}
\caption{The averaged velocities in each $R$ bin for the old RC stars, corresponding to the Figure~\ref{fig:mv_r}.}\label{tab:mv_r2}
\centering
\scriptsize
\begin{tabular}{c|c|c|c|c|c|c|c|c|c}
\hline\hline
$ N $& $\langle R \rangle$ & \vR & \vPhi & \multicolumn{3}{c|}{\VPhi} & \multicolumn{3}{c}{$q$}\\
\hline
 &kpc &\multicolumn{2}{c|}{\kms} &\multicolumn{3}{c|}{\kms}& \multicolumn{3}{c}{} \\
 \hline
 \multicolumn{4}{c|}{Old RC}& \multicolumn{3}{c|}{Old RC}& \multicolumn{3}{c}{Old RC} \\
\hline
\multicolumn{4}{c|}{}& $\phi_0=20\degr$&$\phi_0=30\degr$&$\phi_0=40\degr$& $\phi_0=20\degr$&$\phi_0=30\degr$&$\phi_0=40\degr$\\
\hline
250&7.09&22.19$\pm$4.81&233.23$\pm$2.27&279.51$\pm$13.53&255.06$\pm$5.55&243.36$\pm$3.56&1.18$\pm$0.04&1.12$\pm$0.02&1.10$\pm$0.01\\
250&7.50&-2.79$\pm$4.77&224.04$\pm$1.62&218.38$\pm$6.02&220.63$\pm$3.68&222.38$\pm$2.12&0.97$\pm$0.03&0.98$\pm$0.02&0.99$\pm$0.01\\
250&7.62&-14.26$\pm$5.66&222.27$\pm$1.59&182.16$\pm$21.91&202.96$\pm$8.65&212.76$\pm$4.60&0.77$\pm$0.11&0.88$\pm$0.04&0.92$\pm$0.02\\
1500&7.80&-17.00$\pm$4.68&226.27$\pm$1.17&184.94$\pm$14.43&206.30$\pm$5.44&216.53$\pm$3.03&0.76$\pm$0.07&0.87$\pm$0.02&0.91$\pm$0.01\\
1500&8.11&-2.54$\pm$2.05&225.64$\pm$1.65&220.52$\pm$3.89&222.90$\pm$2.07&224.32$\pm$1.15&0.97$\pm$0.02&0.98$\pm$0.01&0.99$\pm$0.01\\
1500&8.53&-1.89$\pm$1.36&224.05$\pm$1.75&221.05$\pm$1.49&222.53$\pm$1.01&223.40$\pm$0.86&0.98$\pm$0.01&0.99$\pm$0.00&0.99$\pm$0.00\\
1500&8.80&-0.62$\pm$1.26&219.97$\pm$1.77&219.12$\pm$1.04&219.55$\pm$0.90&219.81$\pm$0.85&1.00$\pm$0.00&1.00$\pm$0.00&1.00$\pm$0.00\\
1500&8.94&-2.03$\pm$1.25&222.73$\pm$1.82&220.12$\pm$1.15&221.45$\pm$0.99&222.29$\pm$0.95&0.98$\pm$0.00&0.99$\pm$0.00&0.99$\pm$0.00\\
1500&9.05&-1.80$\pm$0.61&222.82$\pm$1.79&220.53$\pm$1.21&221.70$\pm$1.06&222.44$\pm$1.03&0.99$\pm$0.00&0.99$\pm$0.00&0.99$\pm$0.00\\
1500&9.14&-0.10$\pm$0.60&220.57$\pm$1.82&220.44$\pm$1.16&220.51$\pm$1.08&220.55$\pm$1.05&1.00$\pm$0.00&1.00$\pm$0.00&1.00$\pm$0.00\\
1500&9.24&-0.62$\pm$0.60&223.05$\pm$1.83&222.28$\pm$1.19&222.68$\pm$1.11&222.93$\pm$1.08&1.00$\pm$0.00&1.00$\pm$0.00&1.00$\pm$0.00\\
1500&9.34&3.56$\pm$0.60&224.98$\pm$1.84&229.30$\pm$1.52&227.08$\pm$1.23&225.65$\pm$1.17&0.98$\pm$0.00&0.98$\pm$0.00&0.98$\pm$0.00\\
1500&9.43&1.40$\pm$0.60&220.70$\pm$1.85&222.44$\pm$1.27&221.55$\pm$1.16&220.98$\pm$1.13&0.99$\pm$0.00&0.99$\pm$0.00&0.99$\pm$0.00\\
1500&9.52&1.89$\pm$0.60&215.86$\pm$1.87&218.15$\pm$1.34&216.98$\pm$1.19&216.21$\pm$1.16&0.99$\pm$0.00&0.99$\pm$0.00&0.99$\pm$0.00\\
1500&9.62&2.71$\pm$0.60&222.72$\pm$1.90&226.09$\pm$1.50&224.36$\pm$1.28&223.26$\pm$1.23&0.98$\pm$0.00&0.99$\pm$0.00&0.99$\pm$0.00\\
1500&9.72&2.93$\pm$0.60&214.56$\pm$1.90&218.06$\pm$1.57&216.26$\pm$1.31&215.08$\pm$1.25&0.98$\pm$0.00&0.98$\pm$0.00&0.99$\pm$0.00\\
1500&9.83&3.77$\pm$0.60&219.55$\pm$1.95&224.21$\pm$1.77&221.82$\pm$1.43&220.29$\pm$1.36&0.97$\pm$0.00&0.98$\pm$0.00&0.98$\pm$0.00\\
1500&9.94&2.64$\pm$0.60&215.19$\pm$1.98&218.40$\pm$1.58&216.75$\pm$1.39&215.68$\pm$1.35&0.98$\pm$0.00&0.99$\pm$0.00&0.99$\pm$0.00\\
1500&10.06&5.33$\pm$0.64&221.71$\pm$2.83&228.14$\pm$2.14&224.83$\pm$1.58&222.69$\pm$1.46&0.96$\pm$0.01&0.97$\pm$0.00&0.98$\pm$0.00\\
1500&10.18&3.86$\pm$0.65&218.78$\pm$2.87&223.49$\pm$1.88&221.07$\pm$1.53&219.51$\pm$1.46&0.97$\pm$0.00&0.98$\pm$0.00&0.98$\pm$0.00\\
1500&10.33&4.85$\pm$0.64&218.16$\pm$2.90&224.06$\pm$2.12&221.03$\pm$1.66&219.08$\pm$1.56&0.97$\pm$0.01&0.97$\pm$0.00&0.98$\pm$0.00\\
1500&10.47&5.71$\pm$0.65&214.65$\pm$2.93&221.45$\pm$2.36&217.97$\pm$1.81&215.68$\pm$1.68&0.96$\pm$0.01&0.97$\pm$0.00&0.97$\pm$0.00\\
1500&10.63&4.10$\pm$0.64&216.55$\pm$2.99&221.46$\pm$2.10&218.94$\pm$1.77&217.29$\pm$1.70&0.97$\pm$0.01&0.98$\pm$0.00&0.98$\pm$0.00\\
1500&10.80&4.34$\pm$0.65&221.22$\pm$3.06&226.36$\pm$2.17&223.72$\pm$1.88&221.98$\pm$1.81&0.97$\pm$0.00&0.98$\pm$0.00&0.98$\pm$0.00\\
1500&11.01&5.69$\pm$0.84&217.57$\pm$5.16&224.15$\pm$2.44&220.76$\pm$1.99&218.50$\pm$1.90&0.96$\pm$0.01&0.97$\pm$0.00&0.97$\pm$0.00\\
1500&11.24&5.62$\pm$0.84&229.18$\pm$5.23&235.82$\pm$2.55&232.41$\pm$2.15&230.16$\pm$2.07&0.96$\pm$0.01&0.97$\pm$0.00&0.98$\pm$0.00\\
1500&11.53&6.35$\pm$0.84&223.02$\pm$5.30&230.44$\pm$2.79&226.61$\pm$2.36&224.08$\pm$2.26&0.96$\pm$0.01&0.97$\pm$0.00&0.97$\pm$0.00\\
1500&11.90&5.32$\pm$0.84&226.69$\pm$5.40&232.99$\pm$2.72&229.75$\pm$2.48&227.62$\pm$2.43&0.96$\pm$0.00&0.97$\pm$0.00&0.98$\pm$0.00\\
701&12.29&6.25$\pm$1.76&224.94$\pm$13.00&232.24$\pm$4.14&228.49$\pm$3.99&226.00$\pm$3.95&0.96$\pm$0.01&0.97$\pm$0.00&0.97$\pm$0.00\\
701&12.46&7.64$\pm$1.76&230.78$\pm$13.07&239.56$\pm$4.39&235.04$\pm$4.22&232.01$\pm$4.18&0.95$\pm$0.01&0.96$\pm$0.00&0.97$\pm$0.00\\
701&12.69&6.89$\pm$1.76&238.08$\pm$13.32&245.91$\pm$5.06&241.88$\pm$4.93&239.16$\pm$4.90&0.96$\pm$0.01&0.97$\pm$0.00&0.97$\pm$0.00\\
701&13.00&6.95$\pm$3.25&247.55$\pm$45.33&255.22$\pm$5.49&251.23$\pm$5.36&248.51$\pm$5.33&0.96$\pm$0.01&0.97$\pm$0.00&0.97$\pm$0.00\\
701&13.54&6.75$\pm$3.26&260.66$\pm$45.54&268.18$\pm$7.02&264.26$\pm$6.93&261.62$\pm$6.91&0.96$\pm$0.00&0.97$\pm$0.00&0.97$\pm$0.00\\
\hline
\hline
\end{tabular}
\end{table*}

 %

 \begin{table}[htbp]
\caption{Parameters used in the toy model of the perturbation of the density wave spiral structures}.\label{tab:spiralmodel}
\centering
\begin{tabular}{c|c||c|c}
\hline
\hline
Parameters & Values &Parameters &Values \\
\hline
$m$ & 2 &$\Omega_p$(\kms\ikpc) &$18.6$ \\
\hline
$\mathcal A$ (\%)& $3.09$&$i$ ($\degr$) & $-10.0$ \\
\hline
$\chi_0 ($\degr$)$ & $76.0$ & $M_{disk}$ (\Msun) & $10^{11}$ \\
\hline
$R_d$ (kpc) & 3.5 & $\Sigma_0$ (\Msun$/pc^{2}$) & $1.3\times10^9$ \\
\hline
$A$ (\kms\ikpc) & 14.8 & $B$ (\kms\ikpc) & 18.9  \\
\hline
{$\mathcal{F}_{\nu}^{(1)}$} (old population) & 0.5 &  $\mathcal{F}_{\nu}^{(2)}$ (old population) & 0.5  \\
 \hline
{$\mathcal{F}_{\nu}^{(1)}$} (young population)& 0.8 &  $\mathcal{F}_{\nu}^{(2)}$ (young population)& 0.8  \\
\hline
\hline
\end{tabular}\\
1) $m$ is the number of the spiral arms.
2) $\Omega_p$ is the pattern speed of the spiral structures.
3) $\mathcal A$ is the relative amplitude of the spiral arms.
4) $i$ is the pitch angle of the spiral arms.
5) $\chi_0$ is the phase angle at the position of the Sun.
6) $M_{disk}$ is the total mass of the exponential disk.
7) $R_d$ is the scale length of the exponential disk mass profile.
8) $\Sigma_0$ is the central surface mass density of the disk.
9) $A$ is the first Oort's constant.
10) $B$ is the second Oort's constant.
11) $\mathcal{F}^{(1)}$ is the reduction factor for the radial velocity.
12) $\mathcal{F}^{(2)}$ is the reduction factor for the azimuthal velocity.
\end{table}
 
\clearpage

\end{document}